\documentclass[%
 reprint,
 amsmath,amssymb,
 aps,
]{revtex4-2}

\usepackage{graphicx}
\usepackage{dcolumn}
\usepackage{bm}

\usepackage{subfigure}
\usepackage{color}
\usepackage{soul}

\begin{document}

\preprint{APS/123-QED}

\title{Ferromagnetic resonance response in square artificial spin ice: \\roles of geometry in vertex dynamics and magnetic configurations}


\author{G.A. Gomez-Iriarte}
 \email{grecia@cbpf.br}
\author{D.E. Gonzalez-Chavez}%
\author{R.L. Sommer}%
\author{J.P, Sinnecker}%
\affiliation{Coordination of Condensed Matter, Applied Physics and Nanoscience, Brazilian Center for Physical Research - CBPF, Rio de Janeiro, Brazil.
}%





\begin{abstract}
Artificial spin ice (ASI) represents a class of uniquely structured superlattices comprising geometrically arranged interacting nanomagnets.
This arrangement facilitates the exploration of magnetic frustration phenomena and related spin wave dynamics, contributing to the field of magnonics.
In this work, we present a compressive study of spin wave dynamics in square ASI lattices using micromagnetic simulations, explicitly focusing on ferromagnetic resonance (FMR) and oscillation modes.
We investigate how variations in the aspect ratio of nanomagnets dimensions influence magnetic excitations in both saturated and remanent states.
We analyze the frequency variations of FMR peaks corresponding to edges and bulk oscillation modes in relation to both magnetic configuration and nanomagnets dimensions.
We emphasize the remanent state of square ASI vertex configurations, particularly focusing on their characteristic energy states (Type I, Type II, Type III, and Type IV), while detailing how symmetries influence oscillation modes.
We observe distinct FMR responses originating from vertices, exhibiting unique characteristics for each energy state.
Our results offer a potential method of utilizing the FMR technique to identify energy states in ASI systems.

\end{abstract}

\maketitle

\section{\label{sec:Intro} Introduction}

Artificial spin ice (ASI) comprises an array of interacting nanoscale magnets that mimics atomic frustration in rare earth pyrochlores such as Dy$_{2}$Ti$_{2}$O$_{7}$, Ho$_{2}$Ti$_{2}$O$_{7}$ and Ho$_{2}$Sn$_{2}$O$_{7}$, with spin orientations comparable to the atomic distribution observed in water ice \cite{harris1997,ramirez1994,bramwell2001}. 
Recently, ASI has garnered attention for its resemblance to magnonic crystals, which are metamaterials analogous to photonic crystals, featuring magnetic properties modulated periodically allowing the control of magnons and their band-structure \cite{wang2009}.

Studying the magnetization dynamics of ASI lattices near a magnetic equilibrium configuration can provide valuable insights into the properties of that equilibrium state.
The investigation of spin excitation, or spin waves, can be effectively conducted using experimental methods such as ferromagnetic resonance (FMR) and Brillouin light scattering spectroscopy (BSL) \cite{skjaervo2020,gliga2020}. 
In an FMR experiment, a microwave-frequency driving magnetic field ($H^\mathrm{rf}$) induces a magnetization torque to the sample, initiating a spin precessional motion. 
When the natural frequency of spin waves within the material matches the frequency of the driving magnetic field, the system resonates, giving rise to measurable absorption peaks \cite{kumar2017,rezende2020,kaffash2021}.
Therefore, FMR is a valuable tool to study magnetization dynamics within ASI systems.

ASI lattices have been investigated experimentally, and by micromagnetic simulations in square lattices \cite{wang2006,gliga2013,jungfleisch2016}, kagome lattices \cite{zhang2013,anghinolfi2015,velo2020}, and other geometries \cite{skjaervo2020,lao2018,saha2021}.
In square lattices, FMR techniques have been applied to study ASI in saturated states \cite{jungfleisch2016,li2016, li2017, lendinez2021, ghosh2019, talapatra2020, gartside2021, arora2022, kuchibhotla2023}.
On remanence, the magnetic configurations at the vertices in square ASI lattices determine their energy states, resulting in four different possible energy states identified as Type I, Type II, Type III, and Type IV \cite{wang2006} (see Fig. \ref{fig:types}).
For thick nanomagnets, these magnetic configurations can degenerate due to broken symmetries at the vertex \cite{gliga2015}.
The FMR response on the remanence state has been studied in terms of geometric and material parameters \cite{li2017, lendinez2021, ghosh2019, gartside2021, heyderman2013, iacocca2020, lendinez2021obser}.
While previous studies have observed the FMR response of the four energy states \cite{gliga2013, gliga2015, jungfleisch2016, gartside2021, arora2022}, the FMR response of these energy states remains relatively underexplored as there was no specific focus on characterizing the response of each energy state individually. 

In this work, we present a comprehensive study investigating the FMR response within square ASI lattices through micromagnetic simulations.
Specifically, we thoroughly examine the constituting elements of the ASI lattices  as isolated nanomagnets (nanoislands) and vertex configurations.
Our analysis encompasses the impact of aspect ratio and the direction of the excitation field on both saturated states and remanence states.
We analyze the observed resonant modes and explain their FMR response, including changes in FMR frequencies,  in terms of the involved magnetic energies and symmetries of the system.
We focus mainly on the ASI vertex in remanence and analyze the oscillation modes across energy states.
The observed resonant modes manifest as collective oscillations, originating from localized dipolar fields and magnetization canting at the vertices.
Our results reveal significant variations in the FMR frequencies of these vertex modes across different energy states linked to the magnetic configuration of the vertex.
This highlights the potential utility of utilizing the FMR technique for identifying energy states within ASI lattices.

\section{\label{sec:method} Method}
We simulate and analyze the FMR response of the constituting elements of square ASI lattices, the individual nanoislands, and the vertices of the ASI lattice, where four nanoislands converge.

\begin{figure}[b]
\includegraphics[width=0.5\linewidth]{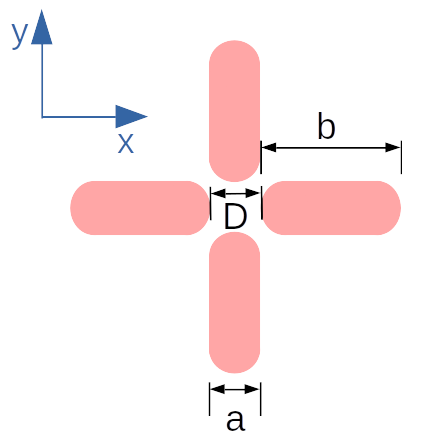}
\caption{Schematic diagram of a square ASI vertex, featuring dimensions of the individual nanoislands: the short axis $a$ remains constant at 80 nm, while the long axis $b$ varies from the set ${160, 240, 400, 800}$ nm. Additionally, the interisland gap $D$ remains fixed at 80 nm.\label{fig:geom_vertex}}
\end{figure}

The nanoislands have a shape as rectangles with rounded sort edges composed of Ni$_{80}$Fe$_{20}$ (Permalloy), featuring a fixed thickness of 10 nm, short axis length $a$ of 80 nm, and rounded edges describing a semicircle. 
The long axis length $b$ is chosen from 160 nm, 240 nm, 400 nm, or 800 nm, resulting in aspect ratios $b$:$a$ of 2:1, 3:1, 5:1, and 10:1, respectively.
For the vertices, we use interisland gap $D$ of 80 nm (see Fig. \ref{fig:geom_vertex}).
The chosen geometries can be easily produced using standard sputtering and e-beam lithography techniques. 
Furthermore, the reduced geometry of the nanoislands and Permalloy's reduced magneto-crystalline anisotropy should ensure the development of nearly uniform magnetization within the nanoislands.

The FMR response was acquired through micromagnetic simulations using the finite-difference discretization based Mumax3 software \cite{vansteenkiste2014}. This software computes time series for the evolution of the reduced magnetization $\vec{m}(\vec{r},t)$ at each position $\vec{r}$ of the discretized grid. 
This computation is performed by integrating the reduced Landau-Lifshitz equation:
\begin{equation}
    \frac{\vec{dm}}{dt}
    = 
    -\frac{\gamma_\mathrm{LL}}{1+\alpha^2} 
    \left( 
    \vec{m} \times \vec{B_\mathrm{eff}} 
    + \alpha \vec{m} \times (\vec{m} \times \vec{B_\mathrm{eff}})
    \right)
    \label{eq:LL}
\end{equation}
with $\gamma_\mathrm{LL}$ been the giromagnetic ratio, $\alpha$ the dimensionless damping parameter and $\vec{B_\mathrm{eff}}$ the effective magnetic field. 
For the effective field contributions we consider demagnetizing, exchange and Zeeman fields.
The material parameters used in the simulations, for permalloy, were: damping coefficient $\alpha=0.008$, saturation magnetization $M_\mathrm{sat} = 8 \times 10^5$ A/m, and exchange stiffness constant $A_\mathrm{ex} = 1.3 \times 10^{-11}$ J/m \cite{buschow2003,vansteenkiste2014,wei2016}.
While, we used a cell size of $5\times5\times5\,\,\mathrm{nm}^3$ for the discretization grid.

The FMR response is calculated employing a pulse field excitation ($H^\mathrm{rf}$) and performing posterior fast Fourier transform (FFT) analysis of the magnetization evolution \cite{kumar2017}. 
Prior to the pulse excitation a stable equilibrium magnetization state was calculated by using the Mumax3 relax function.
Then, the system was driven by an uniform in-plane field $H^\mathrm{rf}$ with the time profile of a $sinc$ function, $H^\mathrm{rf}=h^\mathrm{max} \ sinc(f_\mathrm{c} (t-t_{0}))$, with amplitude $h^\mathrm{max}$ of 5 mT, cut-off frequency $f_\mathrm{c}$ of 20 GHz, and time delay $t_{0}$ of 4 ns. 
During this excitation, the evolution of the magnetization $\vec{m}(\vec{r},t)$ was calculated for a total time of 10 ns and recorded every 10 ps. From this data, the complex valued susceptibility tensor $\chi(\vec{r},f)$ components  were calculated in the frequency space using:
\begin{equation}
    \chi_{\alpha\beta}(\vec{r},f) = 
    \frac{\mathrm{FFT}[\vec{m}_\alpha(\vec{r},t)]} 
         {\mathrm{FFT}[H^\mathrm{rf}_{\beta}(t)]} \quad
\end{equation}
where $\alpha$ and $\beta$ indicate tensor or vector components ($x$, $y$ or $z$) used in the calculation, e.g. $\chi_{zx}$ correspond to the out of plane $z$ magnetization response to an in-plane excitation field in the $x$ direction.
The use of the FFT of the excitation $H^\mathrm{rf}$ in the right size of this equation is crucial to obtain the correct phase of $\chi$.
For direct comparison to broadband FMR experiments is convenient to calculate the power absorbed $P_\mathrm{abs}$ by the system at fixed frequencies. 
This value is proportional to the spatial average of the imaginary part of the $\chi$ component along the direction of  $H^\mathrm{rf}$ in both tensor components, times the frequency, e.g. for $H^\mathrm{rf}$ in the $x$ direction $P_\mathrm{abs}(f) \propto \langle \mathrm{Im}[\chi_{xx}(\vec{r},f)] \rangle f$.
Furthermore, at a resonant frequency $f_r$, the oscillation mode profiles, i.e. the spatial distributions of the amplitude and phase of the oscillating magnetization vector components, are directly proportional to $\chi(\vec{r},f_r)$.

\section{\label{sec:Results} Results and discussion}

\subsection{\label{subsec:macro}Validating the macrospin state in the nanoislands.}

In ASI arrangements, it is essential that each constituting element closely behaves as a macrospin. 
Therefore, the magnetization in the individual nanoislands should be approximately uniform.
To verify this condition, we conducted hysteresis loop simulations with an external field applied along the easy (long) and hard (short) axes of individual nanoislands with different aspect ratios.

\begin{figure} [ht]
        \centering
    \subfigure{\includegraphics[width=0.47\textwidth]{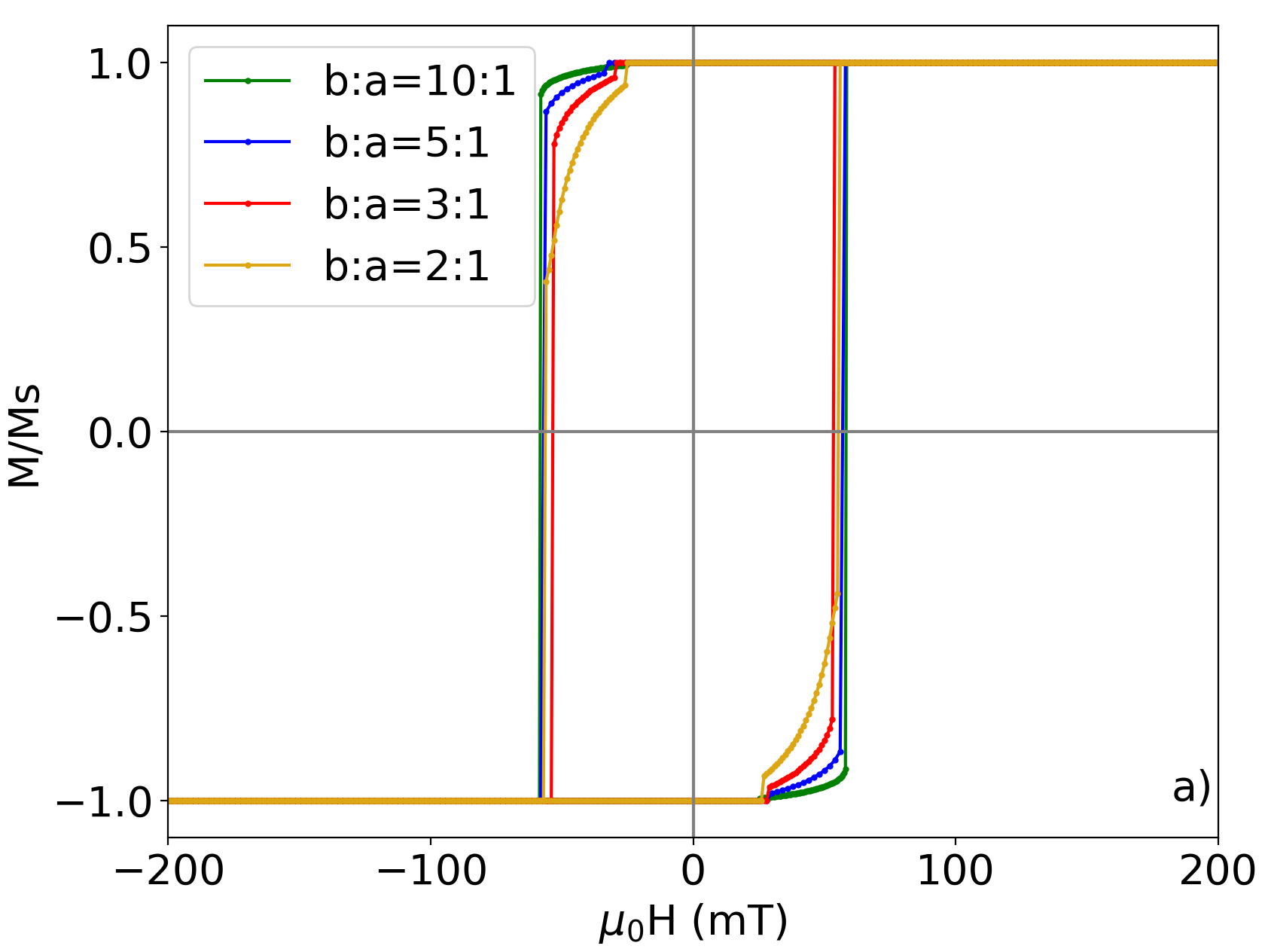}\label{fig:Hist_0}}
    \subfigure{\includegraphics[width=0.46\textwidth]{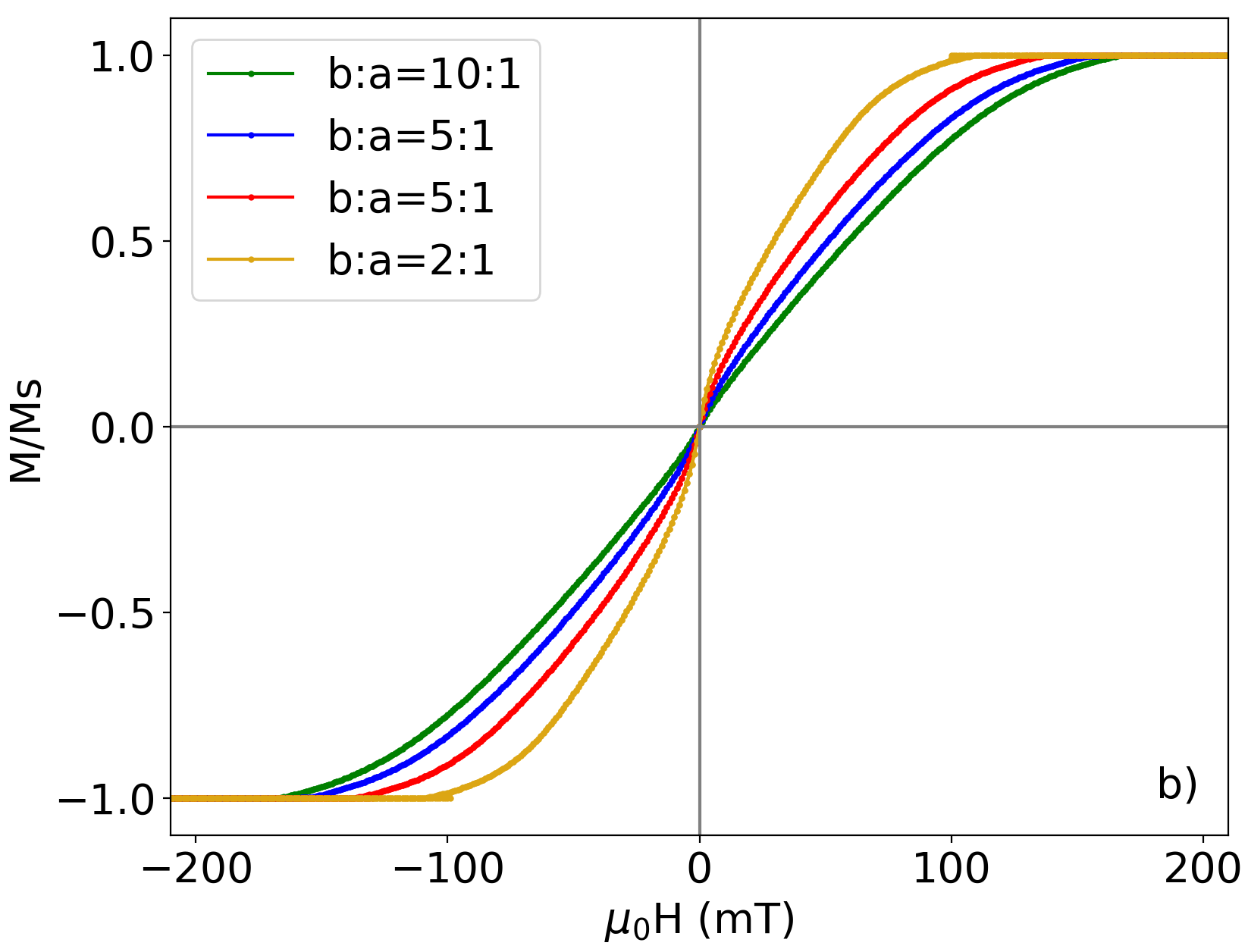}\label{fig:Hist_90}}
    \caption{Hysteresis loops of individual nanoislands by aspect ratio $b$:$a$ = 10:1, 5:1, 3:1 and 2:1, with external magnetic field applied along: a) easy axis, b) hard axis.}\label{fig:hyst}
\end{figure}

Figure \ref{fig:hyst} presents simulated magnetization curves.
The results resemble the Stoner–Wohlfarth \cite{stoner1948,coey2010,guimaraes2009} results expected for the macrospin model.
In this model, the magnetization abruptly inverts at the anisotropy field for the easy axis curves while coherently rotating on the hard axis curves, attaining saturation at the anisotropy field.
In our results, we obtain similar behavior.
Here, the anisotropy field is replaced by the shape anisotropy.
Hysteresis loops for the easy axis (see Fig. \ref{fig:Hist_0}) resulted in square-shaped curves indicating uniform magnetization parallel to the long axis of the nanoislands.
Even if an increment of shape anisotropy is expected for larger aspect ratios, the coercive fields are almost the same for all the simulated aspect ratios.
This indicates that the reversion mechanism does not follow the macrospin behavior; nevertheless, away from the coercive fields, the islands' magnetization behaves as a macrospin.
The magnetization is not uniform only near the coercive fields, especially for lower aspect ratios.
On the other hand, magnetization curves through the hard axis (see Fig. \ref{fig:Hist_90}) reveal a lack of hysteresis, indicating a coherent magnetization rotation. 
As expected, the fields needed for saturation increase with larger shape anisotropy.
In the remanence ($H^\mathrm{bias} = 0$), for all aspect ratios, the equilibrium magnetization is aligned along the long axis of the nanoislands.

\subsection{\label{sec:FMR_nanoisland}FMR response in saturated individual nanoislands.}

\begin{figure*} [ht]
        \centering
        \subfigure{\includegraphics[width=0.45\textwidth]{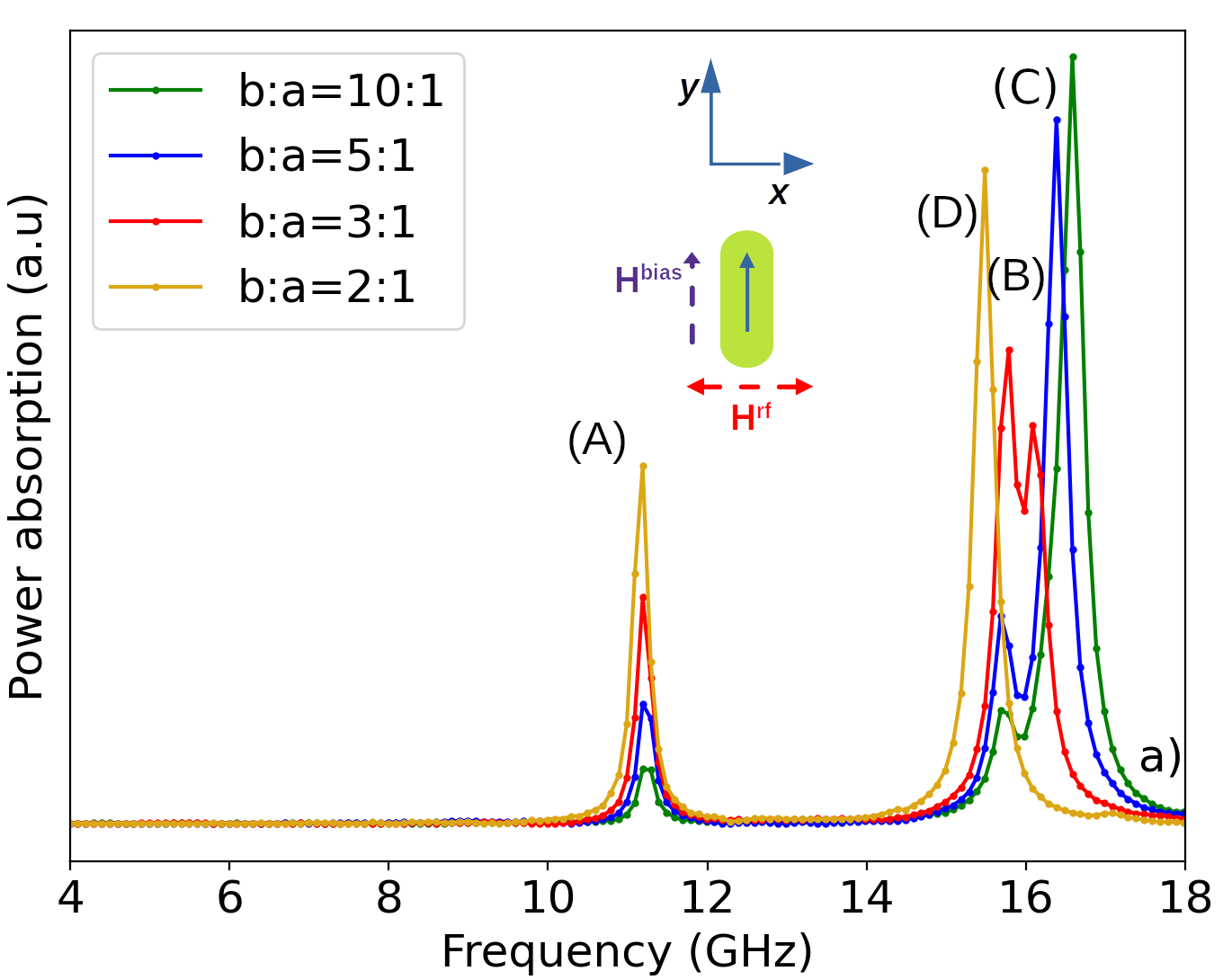}\label{fig:FMR_0}}
        \subfigure{\includegraphics[width=0.45\textwidth]{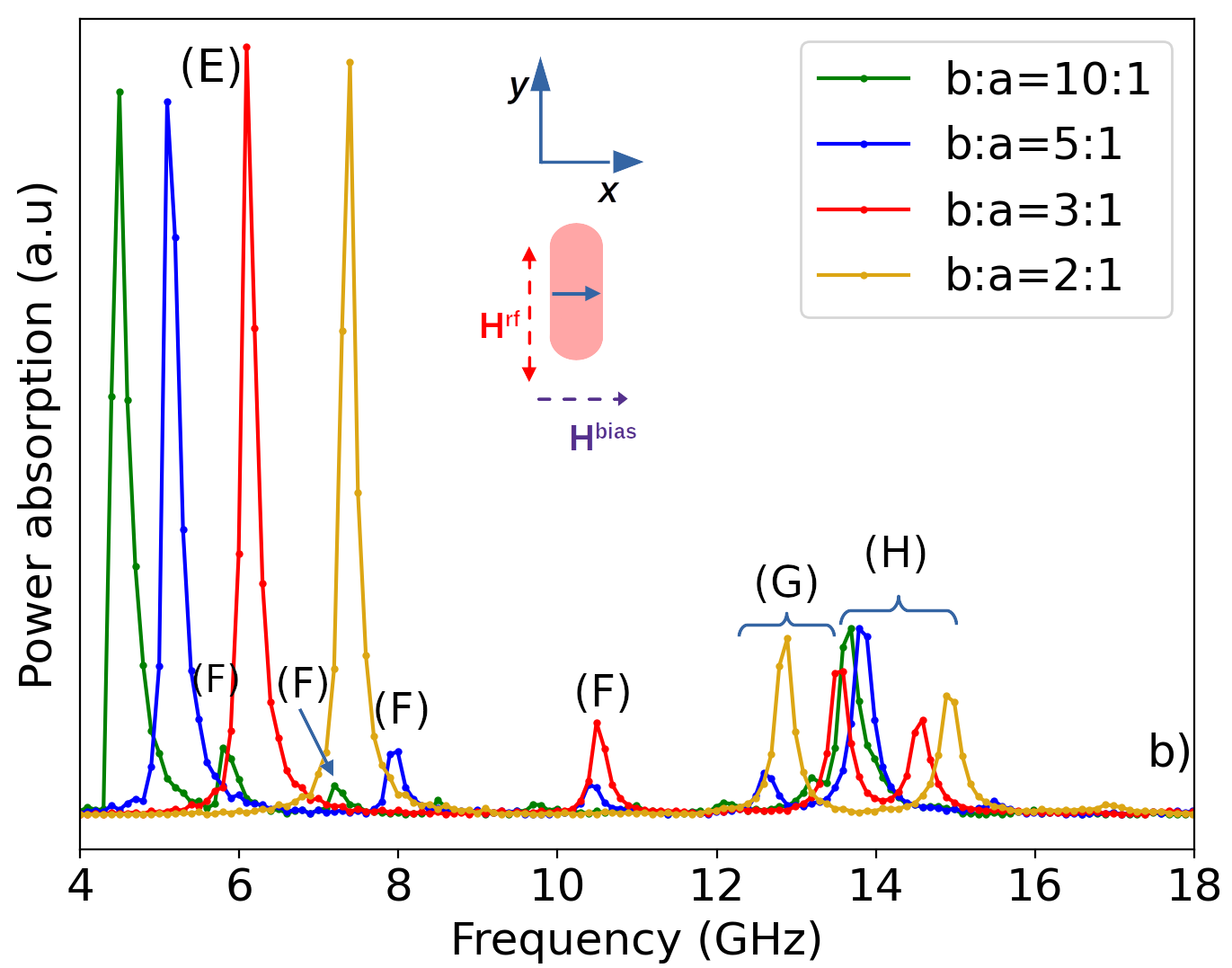}\label{fig:FMR_90}}
        \subfigure{\includegraphics[width=0.9\linewidth]{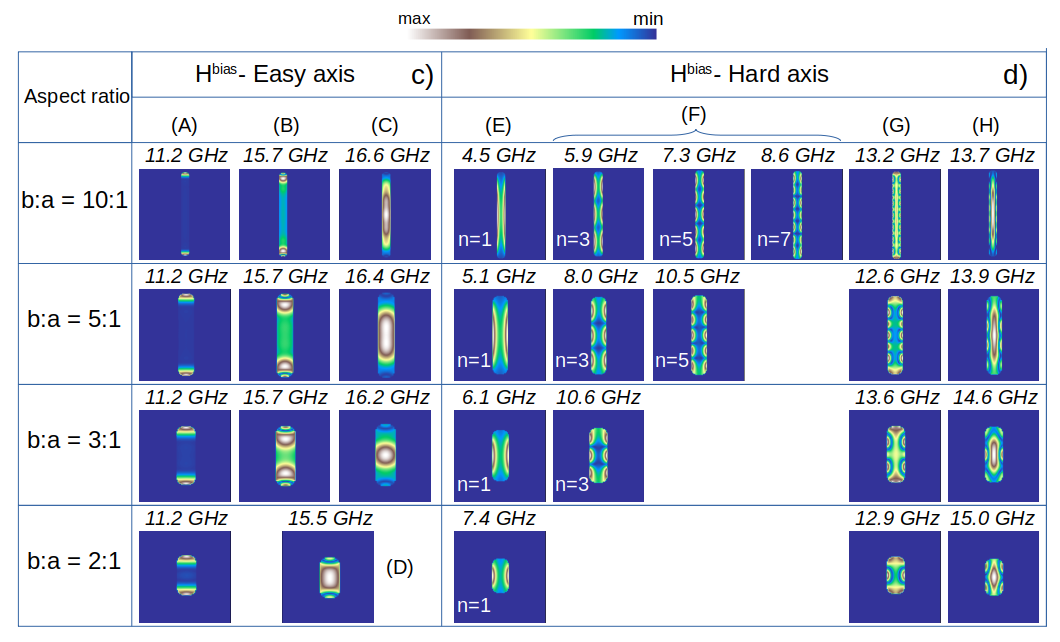}\label{fig:spatial_island}}
        \caption{FMR response from individual islands by aspect ratio: a) FMR spectra with $H^\mathrm{bias}$ along the easy axis, b) FMR spectra with $H^\mathrm{bias}$ along the hard axis, and the respective spatial profiles of the oscillation modes c) and d). } \label{fig:FMRvsF}
   \end{figure*}

Before analyzing the FMR response in the ASI lattice, we study the FMR response of individual nanoislands.
Figure \ref{fig:FMRvsF} displays the simulated FMR spectra and oscillation modes of saturated individual nanoislands as a function of the aspect ratio.
These simulations were conducted with $H^\mathrm{bias}$ = 200 mT, which saturated the sample and applied along the easy or hard axes of the nanoislands.
The in-plane drive field $H^\mathrm{rf}$ is perpendicular to the $H^\mathrm{bias}$ direction. 
The FMR excitation results in standing spin waves with resonance modes localized in the nanoisland edges (edge modes) and resonance modes within the interior of the nanoislands (bulk modes) \cite{arora2022,li2017}.

Figure \ref{fig:FMR_0} illustrates the FMR spectra of individual nanoislands with $H^\mathrm{bias}$ applied along the easy axis. 
The spatial profiles of the oscillation amplitude for the modes at the resonant frequencies for this configuration are presented in figure \ref{fig:FMRvsF}(c). 
The resonance peaks (A) at approximately 11.2 GHz correspond to the main resonance mode from the nanoisland's short edges. 
The frequency of this FMR response is independent of the aspect ratio due to the fixed width of the short axis $a$. 
For aspect ratios 3:1, 5:1, and 10:1, the peak (B) around 15.7 GHz is a higher-order resonance mode that comes from short edges, also maintaining a fixed frequency.
The effect of the aspect ratio becomes evident for peaks (C) related to the bulk modes. 
These exhibit a frequency increment when the aspect ratio increases. 
For the lowest aspect ratio $b$:$a$=2:1, the high order edge mode (B) and the bulk mode (C) hybridize into a single resonance mode (D) observed at 15.5 GHz.

On the other side, figure \ref{fig:FMR_90} depicts the FMR spectra of individual nanoisland with $H_\mathrm{bias}$ field along the hard axis, and the corresponding oscillation modes are shown in figure \ref{fig:FMRvsF}(d). 
The most intense resonance peaks (E) fall between 4 GHz and 7.4 GHz, correlating with the main resonance mode of the long edges of the nanoislands.
The long edges' higher-order resonances (odd harmonic modes) are discernible as smaller peaks (F) between 5.9 GHz and 10.6 GHz. 
Peaks between 12.9 GHz and 13.2 GHz are resonances from a combination of short-edges and long-edges. Finally, the peaks (H) between 13.7 GHz and 15 GHz are bulk modes associated with the nanoisland's interior.
Due to the frequency proximity of peaks (G) and (H), the calculated oscillation profiles are polluted by the mode corresponding to the adjacent peak.
For $H_\mathrm{bias}$ field along the hard axis, the long axis length influences both long-edges modes (E) and bulk modes (H). 
Contrary to the scenario with an easy axis, higher aspect ratios here result in lower frequencies.

The frequency changes in the oscillation modes due to the aspect ratio linked to the shape anisotropy.
The Kittel equation calculated for the macrospin approximation of this system \cite{rezende2020, kittel1947} helps to elucidate these frequency changes. 
Following this equation, the resonant frequency is:
\small
\begin{equation}
    f=\frac{\mu_{0} \gamma}{2\pi} 
    \sqrt{
    \left[ H^\mathrm{bias} + (N_{\perp}^{\alpha} -N_{e})M_\mathrm{sat}\right]
    \left[ H^\mathrm{bias} + (N_{\perp}^{\beta} -N_{e})M_\mathrm{sat}\right]} \label{eq:kittel_sat}
\end{equation}
\normalsize

Here, $\mu_{0}$ denotes the magnetic permeability of free space, while $N_{e}$ stands for the demagnetizing factor aligned with the equilibrium direction.
Additionally, $N_{\perp}^{\alpha}$ and $N_{\perp}^{\beta}$ represent the demagnetizing factors perpendicular to the equilibrium direction, where $\alpha \neq \beta$.

For $H^\mathrm{bias}$ along the easy axis ($y$ direction), we identify $N_{e} = N_{yy}$, $N_{\perp}^{\alpha} = N_{xx}$ and $N_{\perp}^{\beta} = N_{zz}$.
Furthermore, for increasing long axis length, $N_{yy}$ decreases faster than $N_{xx}$ or $N_{zz}$ increases.
For $b$:$a$ =10:1, this results in the highest values for the terms $(N_{xx} -N_{yy})M_\mathrm{sat}$ and $(N_{zz} -N_{yy})M_\mathrm{sat}$, leading to the highest resonance frequency.
Similarly, for $H^\mathrm{bias}$ along the hard axis ($x$ direction), $N_{e} = N_{xx}$, $N_{\perp}^{\alpha} = N_{yy}$ and $N_{\perp}^{\beta} = N_{zz}$.
In this case, the resonant frequency obtained from Eq. \ref{eq:kittel_sat} depends mainly on the term $(N_{yy} -N_{xx})M_\mathrm{sat}$ that is negative and decreases as the aspect ratio increases. Consequently, higher aspect ratios result in lower resonant frequencies.

In order to apply Eq. \ref{eq:kittel_sat} to non-uniform oscillation modes, one must calculate the effective demagnetizing factors. 
These factors depend on the spatial profile of the oscillation modes and are defined by the dimensions of the region where the oscillation takes place \cite{McMichael2006}. 
In particular, for the short edge modes (A), the shape of the regions where these modes are localized does not change with aspect ratio; consequently, no change in the 
effective demagnetizing factors are expected, and the frequencies are the same.
On the other hand, for the long edge modes (E), mode localization results in effective shapes with a larger aspect ratio than the nanoisland itself. 
This results in lower frequencies than bulk modes (H) and lower frequencies for higher aspect ratios.
High-order long-edge modes (F) exhibit frequencies that depend on the number of nodes in the oscillation mode, with the frequency increasing as the node number grows.

\subsection{FMR response as function of bias field in individual nanoislands \label{sec:broadband}}

\begin{figure} [ht]
        \centering
        \subfigure{\includegraphics[width=0.36\textwidth]{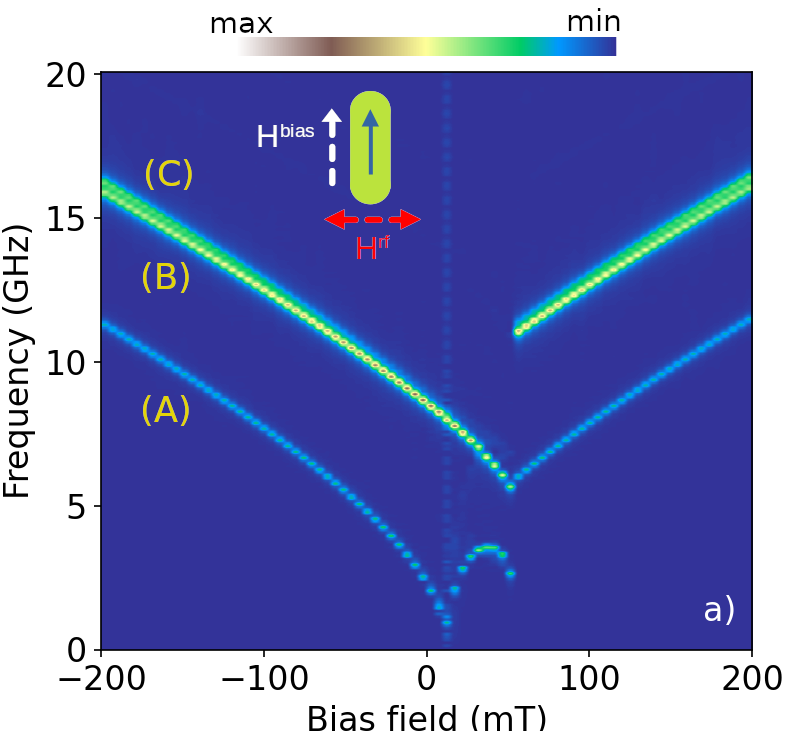}\label{fig:broadband_easy}}
        \subfigure{\includegraphics[width=0.36\textwidth]{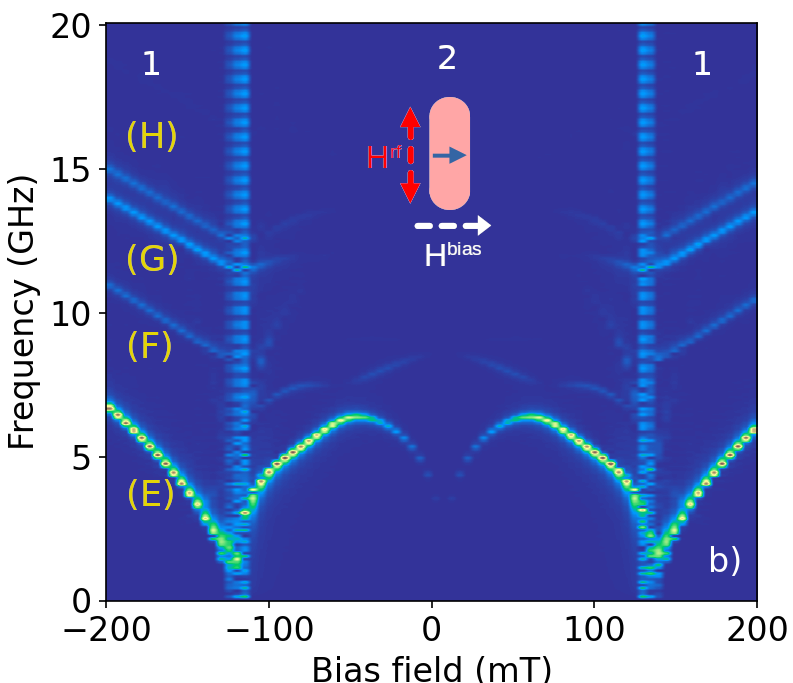}\label{fig:broadband_hard}}
        \subfigure{\includegraphics[width=0.36\textwidth]{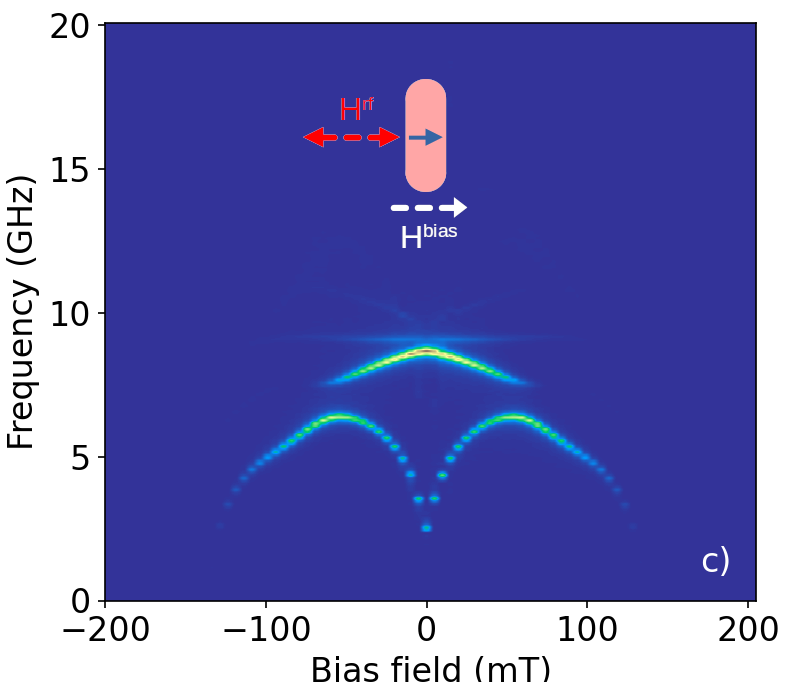}\label{fig:broadband_hardParal}}
        \caption{Broadband FMR response as function of bias field in individual nanoisland for: a) $H^\mathrm{bias}$ applied along easy axis, b) $H^\mathrm{bias}$ along hard axis, c) $H^\mathrm{bias}$ and $H^\mathrm{rf}$ parallel, along the hard axis.\label{FMRvsHbias}}
\end{figure}

Up to this point, our focus has been on analyzing the FMR response of saturated nanoislands. However, when dealing with ASI lattices, the system must be in remanence ($H^\mathrm{bias}=0$).
In this section, we explore the correlation between the oscillation modes of individual nanoislands in both saturated states and remanence.
To investigate this correlation, we conducted an analysis of broadband FMR spectra as a function of the bias field.
Figure \ref{FMRvsHbias} exhibits the simulated broadband FMR spectra on individual nanoislands with an aspect ratio of $b$:$a$ = 3:1. 
This particular aspect ratio approximates the common geometries present in studies of artificial spin ice (ASI)  \cite{goryca2022,porro2019, moller2006,keswani2021,morley2019}.
The bias magnetic field ($H^\mathrm{bias}$) swept from the negative (-200 mT) to positive values (+200 mT).

Figure \ref{fig:broadband_easy} presents the simulated broadband FMR response for $H^\mathrm{bias}$ field applied along the easy axis and $H^\mathrm{rf}$ perpendicular to $H^\mathrm{bias}$.
The continuous resonant branches show the evolution of different oscillation modes of nanoislands. 
The three observed branches correspond to short edge modes (A), high-order short edge modes (B), and bulk modes (C). 
The abrupt changes of resonance branches at $H^\mathrm{bias}$ = +54 mT match the magnetization flip at the coercivity field.
When the  $H^\mathrm{bias}$ advance toward the remanence, (B) and (C) branches approach each other due to the influence of the Zeeman energy over these modes. 
For the aspect ratio $b$:$a$=10:1 and $b$:$a$=5:1, the branch modes (B) and (C) are not affected by Zeeman energy. For $b$:$a$=2:1, these modes disappear, and instead, a hybrid mode branch (D) appears (refer to supplementary information S1).

Figure \ref{fig:broadband_hard} shows the simulated broadband FMR response for $H^\mathrm{bias}$ field applied along the hard axis and $H^\mathrm{rf}$ along the easy axis.
The resonance branches in this figure are segmented into two distinct regions. 
Region 1, with $|H_\mathrm{bias}|>$ 125 mT, corresponds to the saturated state.
In this region, the intense branch is associated with the main long edge mode (E), the next branch  corresponds to the higher-order mode (E), the branch (G) is the combination of short edge mode and long edge mode, and the last is the bulk mode (H).
Region 2, ($|H_\mathrm{bias}|<$ 125 mT), delineates the non-saturated states where the magnetization undergoes a coherent rotation. 
The FMR response vanishes towards $H_\mathrm{bias} = 0$ mT, where $H_\mathrm{rf}$ field is parallel to the magnetization in the remanence state, the spin wave excitation negligible.

To address the complete FMR response in region 2, figure \ref{fig:broadband_hardParal} displays the behavior of broadband FMR when both $H^\mathrm{bias}$ and $H^\mathrm{rf}$ are parallel to the hard axis.
In this configuration, the $H^\mathrm{rf}$ field can excite the resonant modes on the non-saturated states. 
This results in the arc-shaped branches depicted in the figure, which are complementary to the ones observed in Figure \ref{fig:broadband_hard}.
The field span of region 2 changes with an aspect ratio (refer to Supplementary Information S2 and S3), following the variations in the fields needed for saturation, as shown in Fig. \ref{fig:Hist_90}.
We must remark that at exactly $H^\mathrm{bias}=0$, the FMR spectra exhibit peaks at 8.6 GHz for the bulk state and 2.5 GHz for the short edge mode, as depicted in Fig. \ref{fig:broadband_easy} and Fig. \ref{fig:broadband_hardParal}. 
The spectra for both figures coincide because this state represents the remanence.

\subsection{Analyzing FMR Responses in One ASI saturated Vertex}

\begin{figure} [ht]
        \centering
        \subfigure{\includegraphics[width=0.45\textwidth]{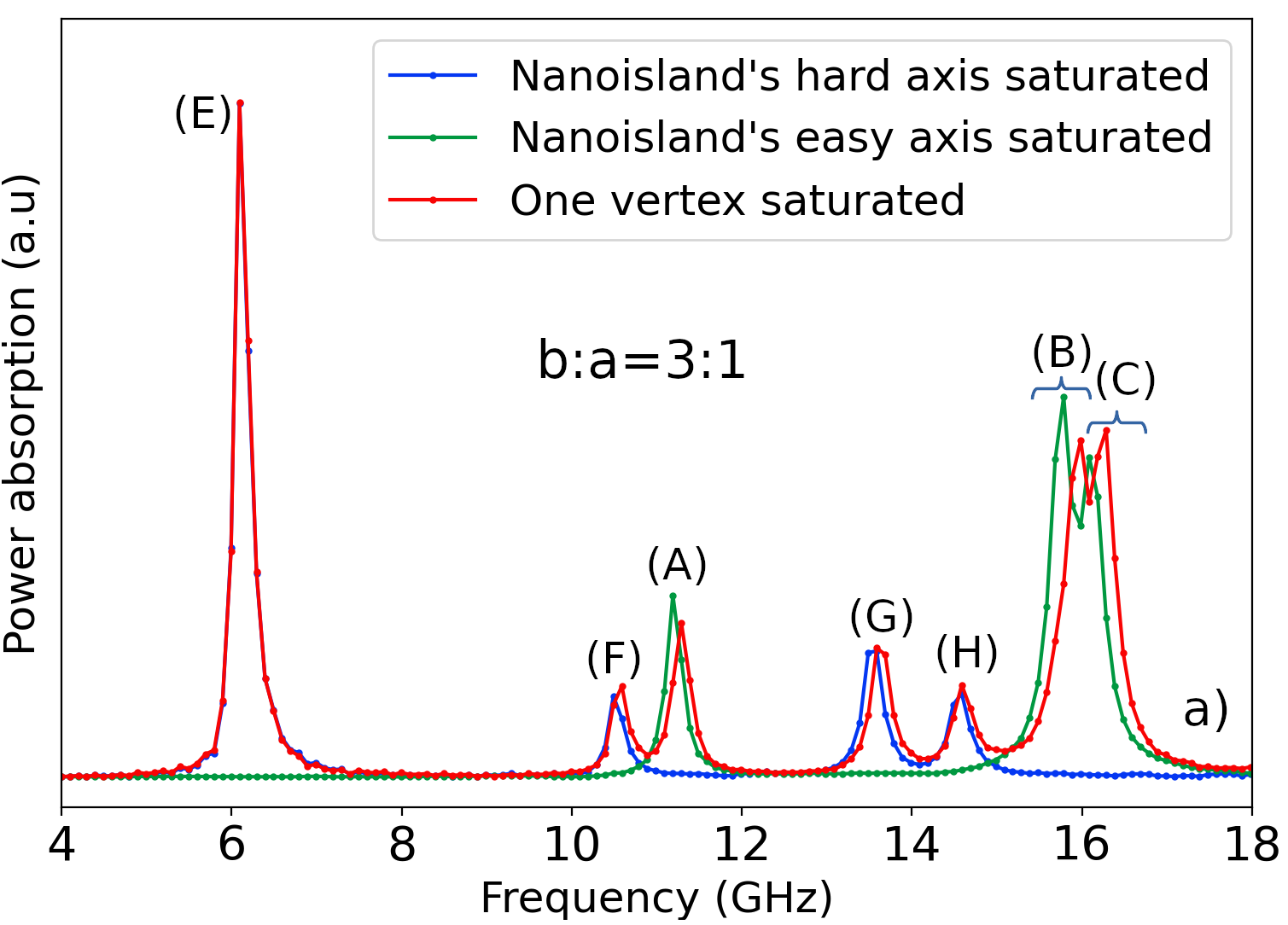}\label{fig:FMR_vertex_sat}}
        \subfigure{\includegraphics[width=0.4\textwidth]{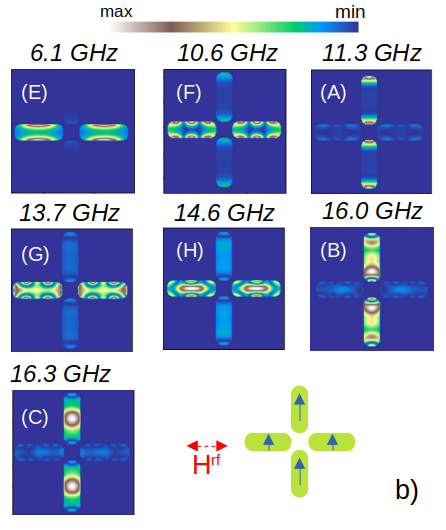}}\label{fig:FMR_vertex_sat_sp}
        \caption{FMR response of one square ASI vertex in the saturated state: a) FMR spectra compared with FMR spectra of individual nanoisland with $H^\mathrm{bias}$ applied along to the hard axis and easy axis. b) Spatial profiles of the oscillation modes. \label{fig:Vertex_sat}}
   \end{figure}

Upon examining the FMR behavior of the essential elements of ASI lattices, the nanoislands, we now analyze ASI lattice interactions at the vertex.
Figure \ref{fig:FMR_vertex_sat} shows FMR spectra of a saturated ASI vertex, compared to the FMR spectra of individual nanoislands, for aspect ratio $b$:$a$= 3:1. 
Peaks associated with the square ASI vertex reveal a superposition of the FMR response of individual nanoisland saturated in the easy or hard axes. 
The spatial mode profiles, showed in fig. \ref{fig:FMR_vertex_sat_sp}, confirm this statement.
This behavior is maintained regardless of the number of simulated vertices in the square ASI lattice. 
Resonance peaks (E), (F), (G), and (H) correspond to FMR response from the horizontal nanoislands, being equivalent to the resonance peaks of individual nanoisland saturated along the hard axis (see fig. \ref{fig:FMR_90}). 
While, Resonance modes (A), (B), and (C) are localized in the vertical islands and correlated to the modes of individual nanoislands saturated along the easy axis (see fig. \ref{fig:FMR_0}). 
However, we note small variations in the FMR frequencies for the same oscillation mode within individual nanoislands and across vertex configurations. 
These variations arise due to the presence of dipolar fields between the nanoislands.

For different aspect ratios (see supplementary information S4), the FMR response preserves the superposition of resonance from horizontal and vertical nanoislands, following the FMR behavior described in section \ref{sec:FMR_nanoisland}.

\subsection{Analyzing FMR Responses in One ASI remanence Vertex \label{sec:reman_vertex}}

\begin{figure}[ht]
    \centering
	\includegraphics[width=3 in]{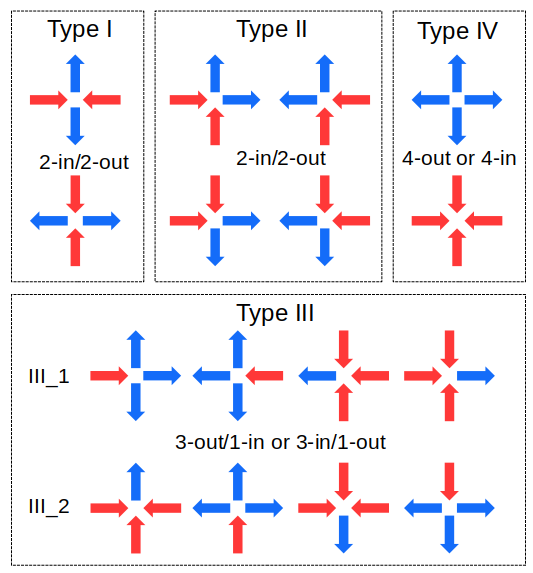}
	\caption{Magnetic configurations at vertex in square ASI lattice. Type III are divided in Type III$\_{1}$ and III$\_{2}$, according to the FMR response.}
	\label{fig:types}
\end{figure}

Artificial Spin Ice (ASI) lattices featuring a square geometry exhibit vertices with four unique energy states, each defined by the magnetic configuration within the vertex \cite{skjaervo2020,gliga2020}, as depicted in Figure \ref{fig:types}. 
The Type I configuration (2-in/2-out), with magnetization in vertical nanoislands opposing each other, is recognized as the ground state.
The various magnetic arrangements for this state result in a doubly degenerate energy state.

Type II (2-in/2-out), with magnetization in vertical nanoislands following each other, is identified as a remanent state due to its ease of attainment after saturation along the diagonal. 
This configuration results in four possible degenerate energy states.
Type III encompasses eight possible degenerate states, featuring magnetic configurations of 3-in/1-out or 1-in/3-out, giving rise to magnetic charge defects analogous to monopoles. 
We classified the Type III state into two subtypes defined by the vertical nanoislands' magnetization orientation:  Type III$\_1$ with opposed magnetization, and Type III$\_2$ with aligned magnetization.
Type IV is the most energetic state, featuring two degenerate states with  4-in or 4-out magnetization, also creating magnetic charges at the vertex. 
In ASI lattices, it is expected that an unbalanced vertex is connected to a vertex with opposite magnetic change, through a string of nanoislands oriented in the appropriate way such that the flux between the two unbalanced vertices is closed \cite{gliga2020, gliga2013}.

In this section we analyzed the FMR response of square ASI vertex for all the energy states. 
The simulations were performed with a $H^\mathrm{bias}=0$ and the $H^\mathrm{rf}$ field applied along the long axis of the horizontal nanoislands. 
Only one magnetic configuration was considered for each energy type, due that the degenerate states presents identical FMR spectra.
The exception is Type III, where Type III$\_1$ and Type III$\_2$ present different FMR response under the simulated conditions.
This discrepancy arises from the magnetization orientation of the vertical nanoislands, which, uniquely in Type III, exhibits two distinct configurations: opposed and aligned.

   \begin{figure} [ht]
        \centering
        \subfigure{\includegraphics[width=0.45\textwidth]{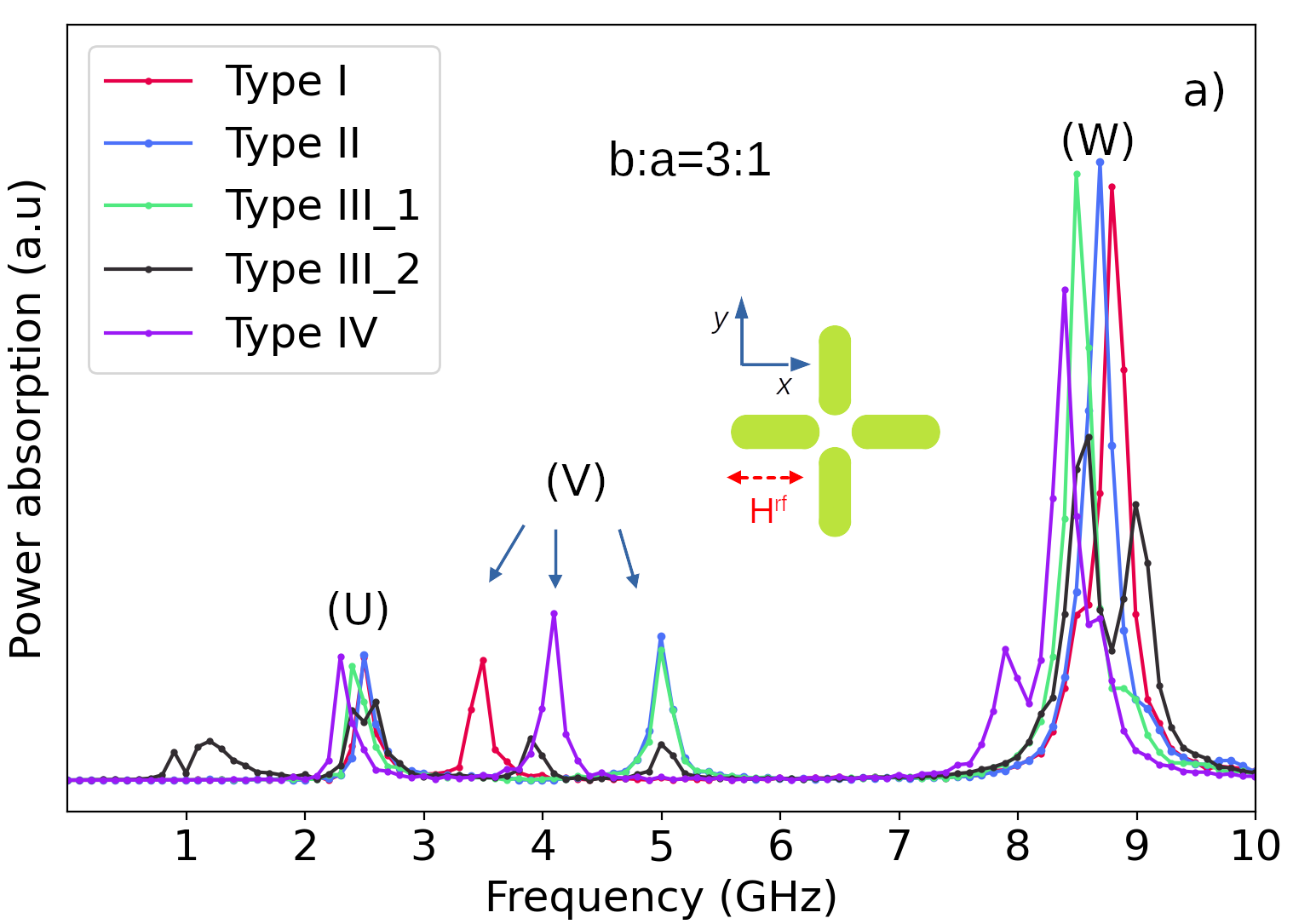}\label{fig:Vertex_rem_1_3_spec}}
        \subfigure{\includegraphics[width=0.43\textwidth]{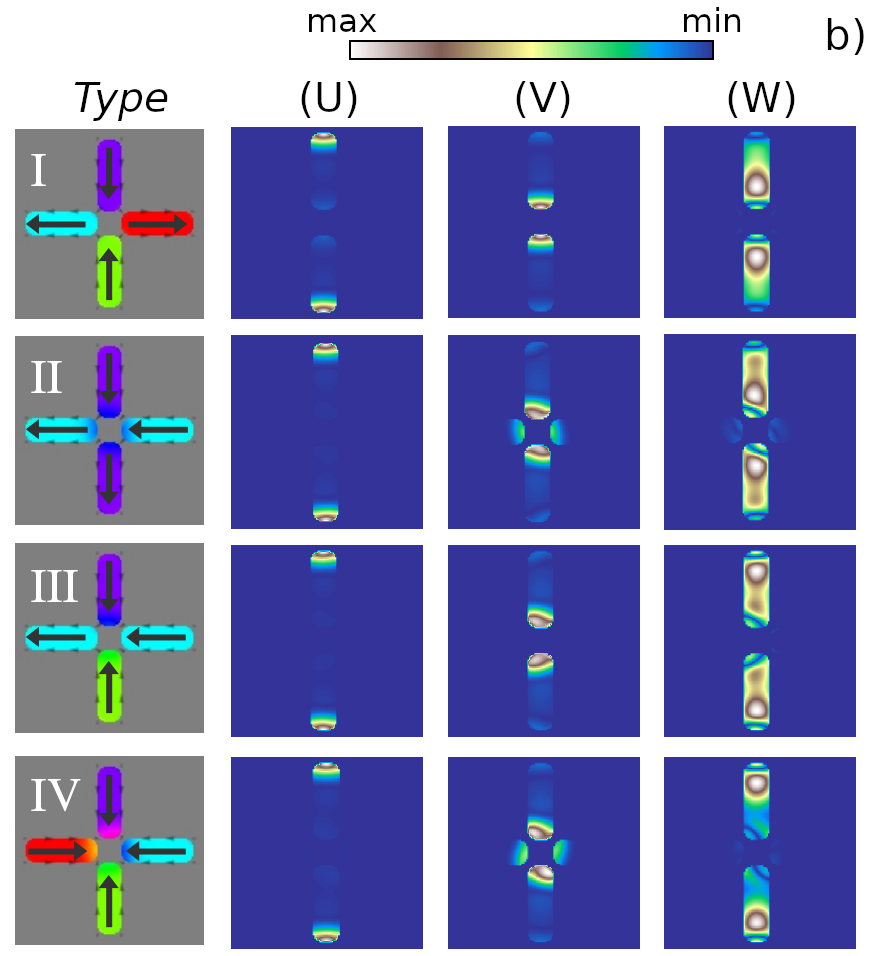}\label{fig:Vertex_rem_1_3_spat}}
        \caption{FMR response of a square ASI vertex in the remanent state for aspect ratio b:a= 3:1:  a) FMR spectra, b) Spatial profile of the oscillation modes.\label{fig:Vertex_rem_1_3}}
   \end{figure}

Figure \ref{fig:Vertex_rem_1_3_spec} displays the FMR spectra of a square ASI vertex with b:a=3:1 aspect ratio for the four energy states.
The selected magnetic configurations are displayed in the left column of Figure \ref{fig:Vertex_rem_1_3_spat}.
Figure \ref{fig:Vertex_rem_1_3_spat} also shows the spatial profile of the resonance modes.
The result reveals three groups of absorption peaks: 
peaks around 2.5 GHz are the response of the oscillations localized on the outer short edge of the vertical nanoislands (U);
peaks between 3.0 GHz and 5.5 GHz correspond to oscillations at the vertices (V);
peaks between 7.8 GHz and 9.4 GHZ are related to the bulk modes (W), and the higher order modes of (U) and (V) oscillations.
Since $H^\mathrm{rf}$ is parallel to the magnetization of the horizontal nanoislands, their oscillation modes are not excited, resulting in a FMR response coming only from the vertical nanoislands modes.
However, horizontal nanoislands can be excited by coupling with vertical nanoislands.
In a complete ASI lattice the FMR signal coming from peaks (U) should be negligible, as it only comes from the short edges of the nanoisland at the edges of the lattice. 
In the following we analyzed the FMR response coming from (V) and (W) oscillation modes.

\subsubsection{FMR response from bulk modes at remanence}

To analyze the FMR response coming from the bulk modes localized in (W) peaks, we approximate each nanoisland as a macrospin.
For one vertex, we have a four-macrospin system, which should result in four unique oscillation modes.
These modes depend on the macrospins' coupling in each particular magnetic configuration.

\begin{figure}[ht]
    \centering
	\includegraphics[width=3.4 in]{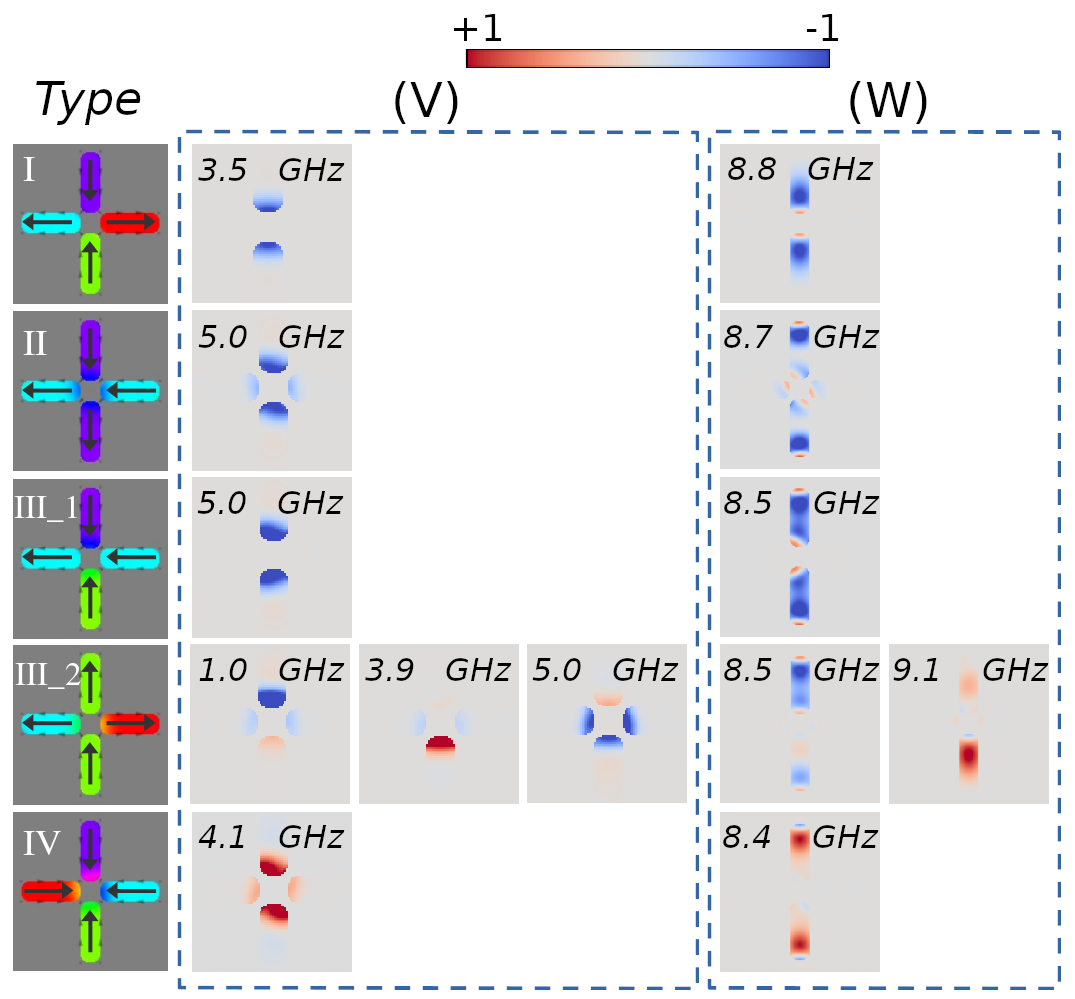}
	\caption{Spatial profiles of the in-plane magnetization oscillations, in quadrature with $H^\mathrm{rf}$, at vertex in a square ASI lattice for aspect ratio 3:1, according to energy state.}
	\label{fig:imag_1_3}
\end{figure}

First, we analyze the bulk modes for a Type III magnetic configuration. 
The results presented for Type III$\_1$ and Type III$\_2$ can be considered to represent the same magnetic configuration, differing only in the orientation of $H^\mathrm{rf}$.
These results account for three distinct FMR peaks, corresponding to three of the four oscillation modes.
To analyse these modes, we make use of the well known results form coupled magnetic layers \cite{Zhang1994, Belmeguenai2008, gonzalez2013, Pervez2022}.
These studies have identified two distinct oscillation modes: in-phase and out-of-phase. 
Furthermore, when the layers differ, one of them governs the oscillation behavior for each mode \cite{gonzalez2013}.
In Type III$\_2$ states, upon examining the magnetic configuration (see Fig. \ref{fig:types}), it becomes evident that the top and bottom vertical nanoislands differs as they possess distinct magnetic energy.
Accordingly, we observe two FMR peaks, with frequencies at 8.5 GHz and 9.1 GHz, with each peak being governed by the oscillations of the bottom and top nanoislands, respectively (refer to Fig. \ref{fig:imag_1_3}).
Moreover, owing to the coupling between the nanoislands, these peaks correspond to the in-phase and out-of-phase oscillation modes. 
However, this distinction may not be immediately discernible in the oscillation profile due to interference from a high-order edge mode. Nevertheless, for an aspect ratio of $b$:$a$=5:1, distinguishing between the in-phase and out-of-phase modes becomes straightforward (See supplementary information S5).
On the other hand, for Type III$\_1$ states, top and bottom vertical nanoisland have exactly the same magnetic energy. 
This results in only one peak at 8.5 GHz that correspond to the in-phase oscillation mode.
In this scenario, the out-of-phase oscillation mode can not be excited by an uniform $H^\mathrm{rf}$, thus it does not appear in the FMR spectrum.
In summary, in a Type III magnetic configuration, $H^\mathrm{rf}$ in one direction can excite two oscillation modes, whereas $H^\mathrm{rf}$ in the perpendicular direction can excite a third mode. 
To excite the remaining mode, a non-uniform excitation field would be required.

For Type I, Type II, and Type IV states, the same analysis employed for Type III is applicable.
However, due to the symmetries in these states, only one FMR peak is observed.
This peak corresponds to the in-phase oscillation of either the vertical or horizontal nano-islands, sharing the same frequency. 
The out-of-phase oscillation modes are not observed as they are not excited by the applied uniform $H^\mathrm{rf}$.

The FMR frequencies for the (W) bulk modes exhibit slight variations depending on the energy state. 
These differences can be explained through Eq. \ref{eq:kittel_sat}.
In this context, the effective demagnetizing factors remain uniform across all energy states, as they are determined solely by geometry rather than magnetic configuration. 
However, individual nanoislands are influenced by the dipolar field generated by adjacent nanoislands. 

\begin{figure} [ht]
        \centering
        \subfigure{\includegraphics[width=0.42\textwidth]{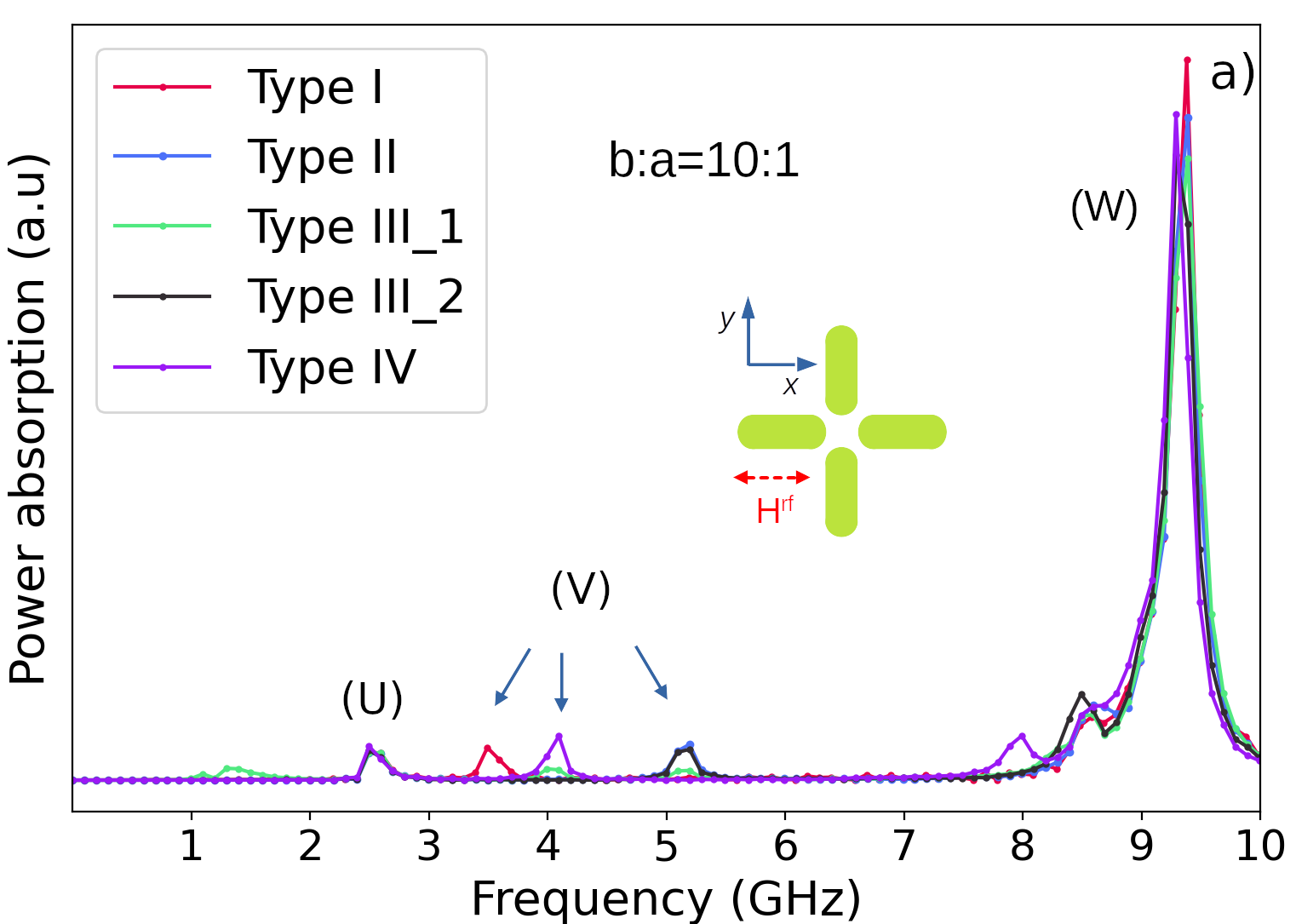}\label{fig:FMRVertex_rem_1_10}}
        \subfigure{\includegraphics[width=0.42\textwidth]{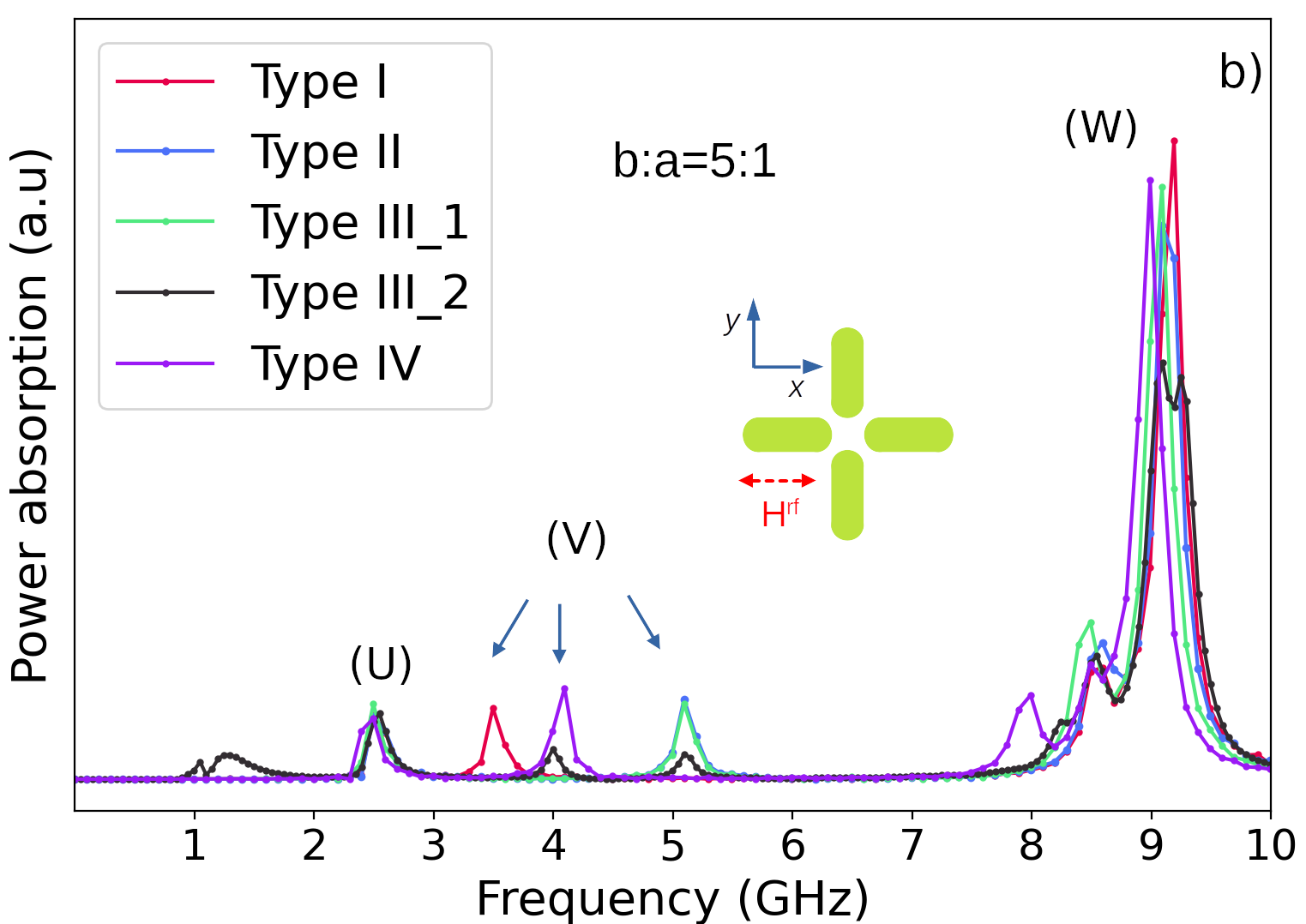}\label{fig:FMRVertex_rem_1_5}}
        \subfigure{\includegraphics[width=0.42\textwidth]{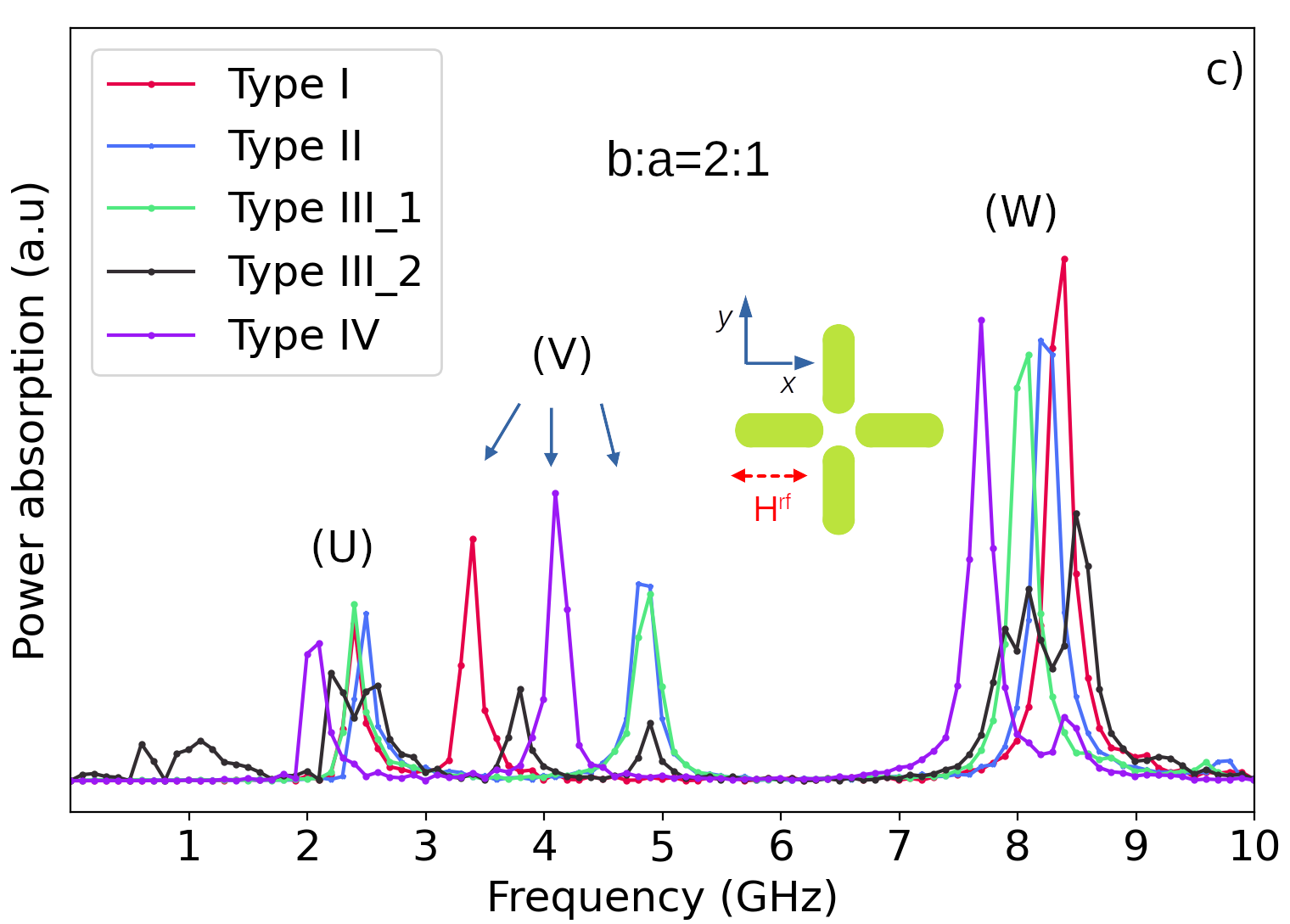}\label{fig:FMRVertex_rem_1_2}}
        \caption{FMR spectra of a square ASI vertex at remanence for different aspect ratios $b$:$a$: a) 10:1, b) 5:1, c) 2:1.\label{fig:vertex_rem}}
\end{figure}

The orientation of this field, relative to the magnetization orientation of the nanoislands, becomes crucial in determining the resonant frequency.
Considering that a nanoisland generates a stronger field over a perpendicular neighbor than over an aligned one, it becomes straightforward to calculate the sign and relative intensities of the total dipolar field for each energy state.
These values are positive for both Type I and Type II configurations. 
Consequently, this leads to higher FMR frequencies (8.7 GHz for Type II and 8.8 GHz for Type I) compared to the FMR frequency of the individual nanoisland (8.6 GHz, see Sec. \ref{sec:broadband}) .
Additionally, the greater frequency for Type I compared to Type II can be attributed to the more intense total dipolar field in Type I configurations.
For Type III$\_1$ configurations on both top and bottom nanoislands, as well as for Type IV configurations, the total dipolar field is negative, resulting in FMR peaks with lower frequencies than the individual nanoisland.
A similar analysis is applicable to Type III$\_2$ configurations; however, in this case, the FMR frequencies are also influenced by the coupling between the top and bottom nanoislands.

As observed in Fig. \ref{fig:vertex_rem}, larger aspect ratios result in smaller frequency variations across energy states in (W) modes. 
We attribute this trend to a decrease in the total dipolar field for such cases.
Additionally, higher aspect ratios yield peaks (W) with greater amplitudes compared to the (U) and (V) peaks. 
At 8.5 GHz, a peak corresponding to a higher-order mode from the outer short edge is also evident. 
Like the (U) peaks, for a fully formed ASI lattice, this peak's amplitude should be negligible.
Furthermore, a high-order mode of the oscillation at the vertex is observed at 7.9 GHz, only for the Type IV state.
For aspect ratio b:a=2:1, distinct energy states enhance the frequency variations for (W) modes. 
However, in this case, the bulk modes hybridize with the high-order short edge modes. 

In summary, (W) bulk modes are influenced by the dipolar field from neighboring nanoislands. 
We successfully explained the minor frequency variations observed for a single vertex under different energy states. 
However, in ASI lattices, the total dipolar field and coupling interactions are considerably more intricate than in the studied case. 
Nevertheless, we anticipate that the expected frequency values should cluster around the FMR frequency of an individual nanoisland.
Experimentally, the FMR response from these modes will likely manifest as a broad peak, making it exceedingly challenging to distinguish the signal arising from different energy states.

\subsubsection{FMR response from vertex modes at remanence}

Contrary to the (W) peaks, (V) peaks present a large frequency variations depending on the energy state, as shown in Fig. \ref{fig:Vertex_rem_1_3_spec}.
Type I and Type IV states have unique FMR frequencies at 3.5 GHz and 4.1 GHz respectively. 
The peak fro Type II and Type III$\_1$ states share the same FMR frequency at 5.0 GHz, while for Type III$\_2$ three peaks ate 1.0 GHz, 3.9 GHz and 5.0 GHz are observed.
The frequency of these (V) peaks do not change with the aspect ratio (See Fig. \ref{fig:vertex_rem}).

The oscillation modes associated with the (V) peaks localize at the short edges, converging at the vertex. 
Unlike the (U) modes, in this instance, the short-edge regions couple to each other, resulting in fundamentally distinct oscillation modes compared to the uncoupled (U) short-edge modes.
The coupling for these modes is governed by the vertex-localized dipolar field and depends on the separation distance $D$ between the nanoislands.
This vertex-localized dipolar field is expect to be large for small $D$, thereby dominating the local energy density at the vertex.
Consequently, it can produce local magnetic textures at the vertex region, canting the magnetization a the edges, effectively minimizing the local energy.
This local magnetic textures however do not significantly impact the macrospin behavior of the nanoisland.
Specific characteristic of the coupling fields and local magnetic textures are determined by the energy state.

In the observed modes showed in figure \ref{fig:imag_1_3}, for Type I and Type III$\_1$ states, at 3.5 GHz and 5.0 GHz respectively, both the top and bottom short edges at the vertex exhibit in-phase oscillations, characterized by the same amplitude. 
Conversely, the short edges from the left and right nanoislands do not oscillate.
On the other hand, in modes Type II, Type III$\_2$ and Type IV we observe oscillation in all short edges at the vertex.
For Type III$\_2$, at 1.0 GHz the oscillation is predominant in the top short edge, with left and right short edges oscillate in phase and bottom short edge oscillating out of phase. 
For the mode at 3.9 GHz, the oscillation is more intense at the bottom short edge, and in this case left and right short edges oscillate out of phase, while the top short edge is in phase. 
Finally at 5.0 GHz the oscillation is in phase and with similar amplitudes on bottom, left and right short edges, while the top short edge is out of phase.
For Type II and Type IV states, the (V) modes at 5.0 GHz and 4.1 GHz respectively, the oscillation is in phase for all the short edges at the vertex.

We noticed that for the short edges for the horizontal nanoislands of Type I and Type III$\_1$ states do not exhibit magnetization canting.
This explains why these states have (V) modes where only the short edges for the vertical nanoislands oscillate.
Moreover, all the other sates show canted magnetization on short edges for the horizontal nanoislands, resulting in oscillation in all four short edges of the vertex.

Similarly to the bulk (W) modes, the (V) modes could be analyzed with the aid of the macrospin approximation, considering macrospins localized at the short edges that converge into the vertex.
However, in this case the coupling fields will be larger and symmetries of the coupling fields will be affected by local magnetic texture at the vertex. 
This result in complex oscillation modes as observed for Type III$\_2$ state.

For an ASI lattice, the (V) peaks could be used to identify the vertex populations based on energy states, analyzing the relative peak amplitudes. 
Unlike the (W) peaks, the (V) peaks do not overlap, therefore they should be experimentally accessible. 
Furthermore, since these peaks arise from localized interactions, the frequency of the FMR response from one vertex remains unaltered. 
Consequently, the frequencies of the (V) peaks are unaffected by the magnetic configuration of neighboring vertices or the number of nanoislands in the ASI lattice.
Even than (V) peaks for Type II and Type III$\_1$ states share the same frequency, it is possible to identify the vertex populations for each type by changing the direction of $H^\mathrm{rf}$.

\section{Conclusion}

Our study systematically explores the broadband FMR response in Artificial Spin Ice (ASI) systems. 
The FMR response manifests as standing spin waves within the nanoislands, which serve as the fundamental building elements of the ASI system.
We have analyzed the impact of the nanoislands aspect ratio on oscillation modes and FMR frequencies, for isolated nanoislands and vertex configurations.
Independent on the nanoislands aspect ratio, we classify the resonant modes into three main groups: short-edge modes, long-edge modes and bulk modes.
Edge modes typically display lower FMR frequencies compared to bulk modes. 
However, higher-order edge modes may reach similar frequencies to bulk modes, leading to a hybridization of FMR responses and subsequent peak overlapping.
We show that the excitation of these modes can be controlled by choosing the direction of the excitation field.

We conducted an extensive analysis of the FMR frequencies of oscillation modes in terms of the effective demagnetization factors of the nanoislands, the dipolar fields within vertex configurations, and symmetries of the magnetic configurations for different energy states. 
This analysis elucidates the observed FMR frequencies and their variations across the different configurations studied.
We shown that FMR frequencies for long-edge modes and bulk modes are heavily influenced by the shape anisotropy and dipolar fields in ASI lattice.
On the other hand, for fixed nanoisland short axis length, FMR frequencies of short-edge modes due to its localization are independent on the aspect ratio.

We investigated vertices in both saturated and remanence states. 
In the saturated state, the FMR response of an ASI lattice is a superposition of oscillation modes from individual nanoislands saturated along either the easy or hard axes, with small variations in the FMR frequencies due to the presence of dipolar fields between the nanoislands.
On the other hand, in remanence states, beside the observed oscillation modes for individual nanoislands, we also observe additional oscillation modes at the vertex.
These new modes manifest as collective short-edge oscillations, originating from localized dipolar fields and magnetization canting at the vertices.
We demonstrate that the FMR frequencies of these vertex modes exhibit significant variations across different energy states associated with the magnetic configuration of the vertex. 
Unlike the FMR response of bulk modes, where FMR peaks for different energy states overlap, the frequency variations in the vertex modes are substantial enough to distinguish them clearly and potentially make them experimentally accessible.
Our studies on individual nanoislands and vertex configurations characterize and provide insights into the FMR response in ASI lattices.

\begin{acknowledgments}
We acknowledge the support of Fundação Carlos Chagas Filho de Amparo à Pesquisa do Estado do Rio de Janeiro - FAPERJ, under grants E-26/201.521/2022 and E-26/200.594/2022.
\end{acknowledgments}

\appendix
\nocite{*}

\bibliography{biblio}

\begin{thebibliography}{46}%
\makeatletter
\providecommand \@ifxundefined [1]{%
 \@ifx{#1\undefined}
}%
\providecommand \@ifnum [1]{%
 \ifnum #1\expandafter \@firstoftwo
 \else \expandafter \@secondoftwo
 \fi
}%
\providecommand \@ifx [1]{%
 \ifx #1\expandafter \@firstoftwo
 \else \expandafter \@secondoftwo
 \fi
}%
\providecommand \natexlab [1]{#1}%
\providecommand \enquote  [1]{``#1''}%
\providecommand \bibnamefont  [1]{#1}%
\providecommand \bibfnamefont [1]{#1}%
\providecommand \citenamefont [1]{#1}%
\providecommand \href@noop [0]{\@secondoftwo}%
\providecommand \href [0]{\begingroup \@sanitize@url \@href}%
\providecommand \@href[1]{\@@startlink{#1}\@@href}%
\providecommand \@@href[1]{\endgroup#1\@@endlink}%
\providecommand \@sanitize@url [0]{\catcode `\\12\catcode `\$12\catcode `\&12\catcode `\#12\catcode `\^12\catcode `\_12\catcode `\%12\relax}%
\providecommand \@@startlink[1]{}%
\providecommand \@@endlink[0]{}%
\providecommand \url  [0]{\begingroup\@sanitize@url \@url }%
\providecommand \@url [1]{\endgroup\@href {#1}{\urlprefix }}%
\providecommand \urlprefix  [0]{URL }%
\providecommand \Eprint [0]{\href }%
\providecommand \doibase [0]{https://doi.org/}%
\providecommand \selectlanguage [0]{\@gobble}%
\providecommand \bibinfo  [0]{\@secondoftwo}%
\providecommand \bibfield  [0]{\@secondoftwo}%
\providecommand \translation [1]{[#1]}%
\providecommand \BibitemOpen [0]{}%
\providecommand \bibitemStop [0]{}%
\providecommand \bibitemNoStop [0]{.\EOS\space}%
\providecommand \EOS [0]{\spacefactor3000\relax}%
\providecommand \BibitemShut  [1]{\csname bibitem#1\endcsname}%
\let\auto@bib@innerbib\@empty
\bibitem [{\citenamefont {Harris}\ \emph {et~al.}(1997)\citenamefont {Harris}, \citenamefont {Bramwell}, \citenamefont {McMorrow}, \citenamefont {Zeiske},\ and\ \citenamefont {Godfrey}}]{harris1997}%
  \BibitemOpen
  \bibfield  {author} {\bibinfo {author} {\bibfnamefont {M.~J.}\ \bibnamefont {Harris}}, \bibinfo {author} {\bibfnamefont {S.~T.}\ \bibnamefont {Bramwell}}, \bibinfo {author} {\bibfnamefont {D.~F.}\ \bibnamefont {McMorrow}}, \bibinfo {author} {\bibfnamefont {T.}~\bibnamefont {Zeiske}},\ and\ \bibinfo {author} {\bibfnamefont {K.~W.}\ \bibnamefont {Godfrey}},\ }\bibfield  {title} {\bibinfo {title} {Geometrical frustration in the ferromagnetic pyrochlore ${\mathrm{ho}}_{2}{\mathrm{ti}}_{2}{O}_{7}$},\ }\href {https://doi.org/10.1103/PhysRevLett.79.2554} {\bibfield  {journal} {\bibinfo  {journal} {Phys. Rev. Lett.}\ }\textbf {\bibinfo {volume} {79}},\ \bibinfo {pages} {2554} (\bibinfo {year} {1997})}\BibitemShut {NoStop}%
\bibitem [{\citenamefont {Ramirez}(1994)}]{ramirez1994}%
  \BibitemOpen
  \bibfield  {author} {\bibinfo {author} {\bibfnamefont {A.}~\bibnamefont {Ramirez}},\ }\bibfield  {title} {\bibinfo {title} {Strongly geometrically frustrated magnets},\ }\href@noop {} {\bibfield  {journal} {\bibinfo  {journal} {Annual Review of Materials Science}\ }\textbf {\bibinfo {volume} {24}},\ \bibinfo {pages} {453} (\bibinfo {year} {1994})}\BibitemShut {NoStop}%
\bibitem [{\citenamefont {Bramwell}\ and\ \citenamefont {Gingras}(2001)}]{bramwell2001}%
  \BibitemOpen
  \bibfield  {author} {\bibinfo {author} {\bibfnamefont {S.~T.}\ \bibnamefont {Bramwell}}\ and\ \bibinfo {author} {\bibfnamefont {M.~J.}\ \bibnamefont {Gingras}},\ }\bibfield  {title} {\bibinfo {title} {Spin ice state in frustrated magnetic pyrochlore materials},\ }\href@noop {} {\bibfield  {journal} {\bibinfo  {journal} {Science}\ }\textbf {\bibinfo {volume} {294}},\ \bibinfo {pages} {1495} (\bibinfo {year} {2001})}\BibitemShut {NoStop}%
\bibitem [{\citenamefont {Wang}\ \emph {et~al.}(2009)\citenamefont {Wang}, \citenamefont {Zhang}, \citenamefont {Lim}, \citenamefont {Ng}, \citenamefont {Kuok}, \citenamefont {Jain},\ and\ \citenamefont {Adeyeye}}]{wang2009}%
  \BibitemOpen
  \bibfield  {author} {\bibinfo {author} {\bibfnamefont {Z.~K.}\ \bibnamefont {Wang}}, \bibinfo {author} {\bibfnamefont {V.~L.}\ \bibnamefont {Zhang}}, \bibinfo {author} {\bibfnamefont {H.~S.}\ \bibnamefont {Lim}}, \bibinfo {author} {\bibfnamefont {S.~C.}\ \bibnamefont {Ng}}, \bibinfo {author} {\bibfnamefont {M.~H.}\ \bibnamefont {Kuok}}, \bibinfo {author} {\bibfnamefont {S.}~\bibnamefont {Jain}},\ and\ \bibinfo {author} {\bibfnamefont {A.~O.}\ \bibnamefont {Adeyeye}},\ }\bibfield  {title} {\bibinfo {title} {Observation of frequency band gaps in a one-dimensional nanostructured magnonic crystal},\ }\href@noop {} {\bibfield  {journal} {\bibinfo  {journal} {Applied Physics Letters}\ }\textbf {\bibinfo {volume} {94}},\ \bibinfo {pages} {083112} (\bibinfo {year} {2009})}\BibitemShut {NoStop}%
\bibitem [{\citenamefont {Skj{\ae}rv{\o}}\ \emph {et~al.}(2020)\citenamefont {Skj{\ae}rv{\o}}, \citenamefont {Marrows}, \citenamefont {Stamps},\ and\ \citenamefont {Heyderman}}]{skjaervo2020}%
  \BibitemOpen
  \bibfield  {author} {\bibinfo {author} {\bibfnamefont {S.~H.}\ \bibnamefont {Skj{\ae}rv{\o}}}, \bibinfo {author} {\bibfnamefont {C.~H.}\ \bibnamefont {Marrows}}, \bibinfo {author} {\bibfnamefont {R.~L.}\ \bibnamefont {Stamps}},\ and\ \bibinfo {author} {\bibfnamefont {L.~J.}\ \bibnamefont {Heyderman}},\ }\bibfield  {title} {\bibinfo {title} {Advances in artificial spin ice},\ }\href@noop {} {\bibfield  {journal} {\bibinfo  {journal} {Nature Reviews Physics}\ }\textbf {\bibinfo {volume} {2}},\ \bibinfo {pages} {13} (\bibinfo {year} {2020})}\BibitemShut {NoStop}%
\bibitem [{\citenamefont {Gliga}\ \emph {et~al.}(2020)\citenamefont {Gliga}, \citenamefont {Iacocca},\ and\ \citenamefont {Heinonen}}]{gliga2020}%
  \BibitemOpen
  \bibfield  {author} {\bibinfo {author} {\bibfnamefont {S.}~\bibnamefont {Gliga}}, \bibinfo {author} {\bibfnamefont {E.}~\bibnamefont {Iacocca}},\ and\ \bibinfo {author} {\bibfnamefont {O.~G.}\ \bibnamefont {Heinonen}},\ }\bibfield  {title} {\bibinfo {title} {Dynamics of reconfigurable artificial spin ice: Toward magnonic functional materials},\ }\href@noop {} {\bibfield  {journal} {\bibinfo  {journal} {APL Materials}\ }\textbf {\bibinfo {volume} {8}} (\bibinfo {year} {2020})}\BibitemShut {NoStop}%
\bibitem [{\citenamefont {Kumar}\ and\ \citenamefont {Adeyeye}(2017)}]{kumar2017}%
  \BibitemOpen
  \bibfield  {author} {\bibinfo {author} {\bibfnamefont {D.}~\bibnamefont {Kumar}}\ and\ \bibinfo {author} {\bibfnamefont {A.~O.}\ \bibnamefont {Adeyeye}},\ }\bibfield  {title} {\bibinfo {title} {Techniques in micromagnetic simulation and analysis},\ }\href@noop {} {\bibfield  {journal} {\bibinfo  {journal} {Journal of Physics D: Applied Physics}\ }\textbf {\bibinfo {volume} {50}},\ \bibinfo {pages} {343001} (\bibinfo {year} {2017})}\BibitemShut {NoStop}%
\bibitem [{\citenamefont {Rezende}(2020)}]{rezende2020}%
  \BibitemOpen
  \bibfield  {author} {\bibinfo {author} {\bibfnamefont {S.~M.}\ \bibnamefont {Rezende}},\ }\href@noop {} {\emph {\bibinfo {title} {Fundamentals of magnonics}}},\ Vol.\ \bibinfo {volume} {969}\ (\bibinfo  {publisher} {Springer},\ \bibinfo {year} {2020})\BibitemShut {NoStop}%
\bibitem [{\citenamefont {Kaffash}\ \emph {et~al.}(2021)\citenamefont {Kaffash}, \citenamefont {Lendinez},\ and\ \citenamefont {Jungfleisch}}]{kaffash2021}%
  \BibitemOpen
  \bibfield  {author} {\bibinfo {author} {\bibfnamefont {M.~T.}\ \bibnamefont {Kaffash}}, \bibinfo {author} {\bibfnamefont {S.}~\bibnamefont {Lendinez}},\ and\ \bibinfo {author} {\bibfnamefont {M.~B.}\ \bibnamefont {Jungfleisch}},\ }\bibfield  {title} {\bibinfo {title} {Nanomagnonics with artificial spin ice},\ }\href@noop {} {\bibfield  {journal} {\bibinfo  {journal} {Physics Letters A}\ }\textbf {\bibinfo {volume} {402}},\ \bibinfo {pages} {127364} (\bibinfo {year} {2021})}\BibitemShut {NoStop}%
\bibitem [{\citenamefont {Wang}\ \emph {et~al.}(2006)\citenamefont {Wang}, \citenamefont {Nisoli}, \citenamefont {Freitas}, \citenamefont {Li}, \citenamefont {McConville}, \citenamefont {Cooley}, \citenamefont {Lund}, \citenamefont {Samarth}, \citenamefont {Leighton}, \citenamefont {Crespi},\ and\ \citenamefont {Schiffer}}]{wang2006}%
  \BibitemOpen
  \bibfield  {author} {\bibinfo {author} {\bibfnamefont {R.~F.}\ \bibnamefont {Wang}}, \bibinfo {author} {\bibfnamefont {C.}~\bibnamefont {Nisoli}}, \bibinfo {author} {\bibfnamefont {R.~S.}\ \bibnamefont {Freitas}}, \bibinfo {author} {\bibfnamefont {J.}~\bibnamefont {Li}}, \bibinfo {author} {\bibfnamefont {W.}~\bibnamefont {McConville}}, \bibinfo {author} {\bibfnamefont {B.~J.}\ \bibnamefont {Cooley}}, \bibinfo {author} {\bibfnamefont {M.~S.}\ \bibnamefont {Lund}}, \bibinfo {author} {\bibfnamefont {N.}~\bibnamefont {Samarth}}, \bibinfo {author} {\bibfnamefont {C.}~\bibnamefont {Leighton}}, \bibinfo {author} {\bibfnamefont {V.~H.}\ \bibnamefont {Crespi}},\ and\ \bibinfo {author} {\bibfnamefont {P.}~\bibnamefont {Schiffer}},\ }\bibfield  {title} {\bibinfo {title} {Artificial ‘spin ice’ in a geometrically frustrated lattice of nanoscale ferromagnetic islands},\ }\href@noop {} {\bibfield  {journal} {\bibinfo  {journal} {Nature}\ }\textbf {\bibinfo {volume} {439}},\ \bibinfo {pages} {303} (\bibinfo {year}
  {2006})}\BibitemShut {NoStop}%
\bibitem [{\citenamefont {Gliga}\ \emph {et~al.}(2013)\citenamefont {Gliga}, \citenamefont {K{\'a}kay}, \citenamefont {Hertel},\ and\ \citenamefont {Heinonen}}]{gliga2013}%
  \BibitemOpen
  \bibfield  {author} {\bibinfo {author} {\bibfnamefont {S.}~\bibnamefont {Gliga}}, \bibinfo {author} {\bibfnamefont {A.}~\bibnamefont {K{\'a}kay}}, \bibinfo {author} {\bibfnamefont {R.}~\bibnamefont {Hertel}},\ and\ \bibinfo {author} {\bibfnamefont {O.~G.}\ \bibnamefont {Heinonen}},\ }\bibfield  {title} {\bibinfo {title} {Spectral analysis of topological defects in an artificial spin-ice lattice},\ }\href@noop {} {\bibfield  {journal} {\bibinfo  {journal} {Physical review letters}\ }\textbf {\bibinfo {volume} {110}},\ \bibinfo {pages} {117205} (\bibinfo {year} {2013})}\BibitemShut {NoStop}%
\bibitem [{\citenamefont {Jungfleisch}\ \emph {et~al.}(2016)\citenamefont {Jungfleisch}, \citenamefont {Zhang}, \citenamefont {Iacocca}, \citenamefont {Sklenar}, \citenamefont {Ding}, \citenamefont {Jiang}, \citenamefont {Zhang}, \citenamefont {Pearson}, \citenamefont {Novosad}, \citenamefont {Ketterson}, \citenamefont {Heinonen},\ and\ \citenamefont {Hoffmann}}]{jungfleisch2016}%
  \BibitemOpen
  \bibfield  {author} {\bibinfo {author} {\bibfnamefont {M.~B.}\ \bibnamefont {Jungfleisch}}, \bibinfo {author} {\bibfnamefont {W.}~\bibnamefont {Zhang}}, \bibinfo {author} {\bibfnamefont {E.}~\bibnamefont {Iacocca}}, \bibinfo {author} {\bibfnamefont {J.}~\bibnamefont {Sklenar}}, \bibinfo {author} {\bibfnamefont {J.}~\bibnamefont {Ding}}, \bibinfo {author} {\bibfnamefont {W.}~\bibnamefont {Jiang}}, \bibinfo {author} {\bibfnamefont {S.}~\bibnamefont {Zhang}}, \bibinfo {author} {\bibfnamefont {J.~E.}\ \bibnamefont {Pearson}}, \bibinfo {author} {\bibfnamefont {V.}~\bibnamefont {Novosad}}, \bibinfo {author} {\bibfnamefont {J.~B.}\ \bibnamefont {Ketterson}}, \bibinfo {author} {\bibfnamefont {O.}~\bibnamefont {Heinonen}},\ and\ \bibinfo {author} {\bibfnamefont {A.}~\bibnamefont {Hoffmann}},\ }\bibfield  {title} {\bibinfo {title} {Dynamic response of an artificial square spin ice},\ }\href {https://doi.org/10.1103/PhysRevB.93.100401} {\bibfield  {journal} {\bibinfo  {journal} {Phys. Rev. B}\ }\textbf {\bibinfo
  {volume} {93}},\ \bibinfo {pages} {100401} (\bibinfo {year} {2016})}\BibitemShut {NoStop}%
\bibitem [{\citenamefont {Zhang}\ \emph {et~al.}(2013)\citenamefont {Zhang}, \citenamefont {Gilbert}, \citenamefont {Nisoli}, \citenamefont {Chern}, \citenamefont {Erickson}, \citenamefont {O’brien}, \citenamefont {Leighton}, \citenamefont {Lammert}, \citenamefont {Crespi},\ and\ \citenamefont {Schiffer}}]{zhang2013}%
  \BibitemOpen
  \bibfield  {author} {\bibinfo {author} {\bibfnamefont {S.}~\bibnamefont {Zhang}}, \bibinfo {author} {\bibfnamefont {I.}~\bibnamefont {Gilbert}}, \bibinfo {author} {\bibfnamefont {C.}~\bibnamefont {Nisoli}}, \bibinfo {author} {\bibfnamefont {G.-W.}\ \bibnamefont {Chern}}, \bibinfo {author} {\bibfnamefont {M.~J.}\ \bibnamefont {Erickson}}, \bibinfo {author} {\bibfnamefont {L.}~\bibnamefont {O’brien}}, \bibinfo {author} {\bibfnamefont {C.}~\bibnamefont {Leighton}}, \bibinfo {author} {\bibfnamefont {P.~E.}\ \bibnamefont {Lammert}}, \bibinfo {author} {\bibfnamefont {V.~H.}\ \bibnamefont {Crespi}},\ and\ \bibinfo {author} {\bibfnamefont {P.}~\bibnamefont {Schiffer}},\ }\bibfield  {title} {\bibinfo {title} {Crystallites of magnetic charges in artificial spin ice},\ }\href@noop {} {\bibfield  {journal} {\bibinfo  {journal} {Nature}\ }\textbf {\bibinfo {volume} {500}},\ \bibinfo {pages} {553} (\bibinfo {year} {2013})}\BibitemShut {NoStop}%
\bibitem [{\citenamefont {Anghinolfi}\ \emph {et~al.}(2015)\citenamefont {Anghinolfi}, \citenamefont {Luetkens}, \citenamefont {Perron}, \citenamefont {Flokstra}, \citenamefont {Sendetskyi}, \citenamefont {Suter}, \citenamefont {Prokscha}, \citenamefont {Derlet}, \citenamefont {Lee},\ and\ \citenamefont {Heyderman}}]{anghinolfi2015}%
  \BibitemOpen
  \bibfield  {author} {\bibinfo {author} {\bibfnamefont {L.}~\bibnamefont {Anghinolfi}}, \bibinfo {author} {\bibfnamefont {H.}~\bibnamefont {Luetkens}}, \bibinfo {author} {\bibfnamefont {J.}~\bibnamefont {Perron}}, \bibinfo {author} {\bibfnamefont {M.~G.}\ \bibnamefont {Flokstra}}, \bibinfo {author} {\bibfnamefont {O.}~\bibnamefont {Sendetskyi}}, \bibinfo {author} {\bibfnamefont {A.}~\bibnamefont {Suter}}, \bibinfo {author} {\bibfnamefont {T.}~\bibnamefont {Prokscha}}, \bibinfo {author} {\bibfnamefont {P.~M.}\ \bibnamefont {Derlet}}, \bibinfo {author} {\bibfnamefont {S.~L.}\ \bibnamefont {Lee}},\ and\ \bibinfo {author} {\bibfnamefont {L.~J.}\ \bibnamefont {Heyderman}},\ }\bibfield  {title} {\bibinfo {title} {Thermodynamic phase transitions in a frustrated magnetic metamaterial},\ }\href@noop {} {\bibfield  {journal} {\bibinfo  {journal} {Nature communications}\ }\textbf {\bibinfo {volume} {6}},\ \bibinfo {pages} {8278} (\bibinfo {year} {2015})}\BibitemShut {NoStop}%
\bibitem [{\citenamefont {Ferreira~Velo}\ \emph {et~al.}(2020)\citenamefont {Ferreira~Velo}, \citenamefont {Malvezzi~Cecchi},\ and\ \citenamefont {Pirota}}]{velo2020}%
  \BibitemOpen
  \bibfield  {author} {\bibinfo {author} {\bibfnamefont {M.}~\bibnamefont {Ferreira~Velo}}, \bibinfo {author} {\bibfnamefont {B.}~\bibnamefont {Malvezzi~Cecchi}},\ and\ \bibinfo {author} {\bibfnamefont {K.~R.}\ \bibnamefont {Pirota}},\ }\bibfield  {title} {\bibinfo {title} {Micromagnetic simulations of magnetization reversal in kagome artificial spin ice},\ }\href {https://doi.org/10.1103/PhysRevB.102.224420} {\bibfield  {journal} {\bibinfo  {journal} {Phys. Rev. B}\ }\textbf {\bibinfo {volume} {102}},\ \bibinfo {pages} {224420} (\bibinfo {year} {2020})}\BibitemShut {NoStop}%
\bibitem [{\citenamefont {Lao}\ \emph {et~al.}(2018)\citenamefont {Lao}, \citenamefont {Caravelli}, \citenamefont {Sheikh}, \citenamefont {Sklenar}, \citenamefont {Gardeazabal}, \citenamefont {Watts}, \citenamefont {Albrecht}, \citenamefont {Scholl}, \citenamefont {Dahmen}, \citenamefont {Nisoli} \emph {et~al.}}]{lao2018}%
  \BibitemOpen
  \bibfield  {author} {\bibinfo {author} {\bibfnamefont {Y.}~\bibnamefont {Lao}}, \bibinfo {author} {\bibfnamefont {F.}~\bibnamefont {Caravelli}}, \bibinfo {author} {\bibfnamefont {M.}~\bibnamefont {Sheikh}}, \bibinfo {author} {\bibfnamefont {J.}~\bibnamefont {Sklenar}}, \bibinfo {author} {\bibfnamefont {D.}~\bibnamefont {Gardeazabal}}, \bibinfo {author} {\bibfnamefont {J.~D.}\ \bibnamefont {Watts}}, \bibinfo {author} {\bibfnamefont {A.~M.}\ \bibnamefont {Albrecht}}, \bibinfo {author} {\bibfnamefont {A.}~\bibnamefont {Scholl}}, \bibinfo {author} {\bibfnamefont {K.}~\bibnamefont {Dahmen}}, \bibinfo {author} {\bibfnamefont {C.}~\bibnamefont {Nisoli}}, \emph {et~al.},\ }\bibfield  {title} {\bibinfo {title} {Classical topological order in the kinetics of artificial spin ice},\ }\href@noop {} {\bibfield  {journal} {\bibinfo  {journal} {Nature Physics}\ }\textbf {\bibinfo {volume} {14}},\ \bibinfo {pages} {723} (\bibinfo {year} {2018})}\BibitemShut {NoStop}%
\bibitem [{\citenamefont {Saha}\ \emph {et~al.}(2021)\citenamefont {Saha}, \citenamefont {Zhou}, \citenamefont {Hofhuis}, \citenamefont {K{\'a}kay}, \citenamefont {Scagnoli}, \citenamefont {Heyderman},\ and\ \citenamefont {Gliga}}]{saha2021}%
  \BibitemOpen
  \bibfield  {author} {\bibinfo {author} {\bibfnamefont {S.}~\bibnamefont {Saha}}, \bibinfo {author} {\bibfnamefont {J.}~\bibnamefont {Zhou}}, \bibinfo {author} {\bibfnamefont {K.}~\bibnamefont {Hofhuis}}, \bibinfo {author} {\bibfnamefont {A.}~\bibnamefont {K{\'a}kay}}, \bibinfo {author} {\bibfnamefont {V.}~\bibnamefont {Scagnoli}}, \bibinfo {author} {\bibfnamefont {L.~J.}\ \bibnamefont {Heyderman}},\ and\ \bibinfo {author} {\bibfnamefont {S.}~\bibnamefont {Gliga}},\ }\bibfield  {title} {\bibinfo {title} {Spin-wave dynamics and symmetry breaking in an artificial spin ice},\ }\href@noop {} {\bibfield  {journal} {\bibinfo  {journal} {Nano Letters}\ }\textbf {\bibinfo {volume} {21}},\ \bibinfo {pages} {2382} (\bibinfo {year} {2021})}\BibitemShut {NoStop}%
\bibitem [{\citenamefont {Li}\ \emph {et~al.}(2016)\citenamefont {Li}, \citenamefont {Gubbiotti}, \citenamefont {Casoli}, \citenamefont {Gon{\c{c}}alves}, \citenamefont {Morley}, \citenamefont {Rosamond}, \citenamefont {Linfield}, \citenamefont {Marrows}, \citenamefont {McVitie},\ and\ \citenamefont {Stamps}}]{li2016}%
  \BibitemOpen
  \bibfield  {author} {\bibinfo {author} {\bibfnamefont {Y.}~\bibnamefont {Li}}, \bibinfo {author} {\bibfnamefont {G.}~\bibnamefont {Gubbiotti}}, \bibinfo {author} {\bibfnamefont {F.}~\bibnamefont {Casoli}}, \bibinfo {author} {\bibfnamefont {F.~J.~T.}\ \bibnamefont {Gon{\c{c}}alves}}, \bibinfo {author} {\bibfnamefont {S.~A.}\ \bibnamefont {Morley}}, \bibinfo {author} {\bibfnamefont {M.~C.}\ \bibnamefont {Rosamond}}, \bibinfo {author} {\bibfnamefont {E.~H.}\ \bibnamefont {Linfield}}, \bibinfo {author} {\bibfnamefont {C.~H.}\ \bibnamefont {Marrows}}, \bibinfo {author} {\bibfnamefont {S.}~\bibnamefont {McVitie}},\ and\ \bibinfo {author} {\bibfnamefont {R.~L.}\ \bibnamefont {Stamps}},\ }\bibfield  {title} {\bibinfo {title} {Brillouin light scattering study of magnetic-element normal modes in a square artificial spin ice geometry},\ }\href@noop {} {\bibfield  {journal} {\bibinfo  {journal} {Journal of Physics D: Applied Physics}\ }\textbf {\bibinfo {volume} {50}},\ \bibinfo {pages} {015003} (\bibinfo {year}
  {2016})}\BibitemShut {NoStop}%
\bibitem [{\citenamefont {Li}\ \emph {et~al.}(2017)\citenamefont {Li}, \citenamefont {Gubbiotti}, \citenamefont {Casoli}, \citenamefont {Morley}, \citenamefont {Goncalves}, \citenamefont {Rosamond}, \citenamefont {Linfield}, \citenamefont {Marrows}, \citenamefont {McVitie},\ and\ \citenamefont {Stamps}}]{li2017}%
  \BibitemOpen
  \bibfield  {author} {\bibinfo {author} {\bibfnamefont {Y.}~\bibnamefont {Li}}, \bibinfo {author} {\bibfnamefont {G.}~\bibnamefont {Gubbiotti}}, \bibinfo {author} {\bibfnamefont {F.}~\bibnamefont {Casoli}}, \bibinfo {author} {\bibfnamefont {S.~A.}\ \bibnamefont {Morley}}, \bibinfo {author} {\bibfnamefont {F.~J.~T.}\ \bibnamefont {Goncalves}}, \bibinfo {author} {\bibfnamefont {M.~C.}\ \bibnamefont {Rosamond}}, \bibinfo {author} {\bibfnamefont {E.~H.}\ \bibnamefont {Linfield}}, \bibinfo {author} {\bibfnamefont {C.~H.}\ \bibnamefont {Marrows}}, \bibinfo {author} {\bibfnamefont {S.}~\bibnamefont {McVitie}},\ and\ \bibinfo {author} {\bibfnamefont {R.~L.}\ \bibnamefont {Stamps}},\ }\bibfield  {title} {\bibinfo {title} {Thickness dependence of spin wave excitations in an artificial square spin ice-like geometry},\ }\href@noop {} {\bibfield  {journal} {\bibinfo  {journal} {Journal of Applied Physics}\ }\textbf {\bibinfo {volume} {121}} (\bibinfo {year} {2017})}\BibitemShut {NoStop}%
\bibitem [{\citenamefont {Lendinez}\ \emph {et~al.}(2021{\natexlab{a}})\citenamefont {Lendinez}, \citenamefont {Kaffash},\ and\ \citenamefont {Jungfleisch}}]{lendinez2021}%
  \BibitemOpen
  \bibfield  {author} {\bibinfo {author} {\bibfnamefont {S.}~\bibnamefont {Lendinez}}, \bibinfo {author} {\bibfnamefont {M.~T.}\ \bibnamefont {Kaffash}},\ and\ \bibinfo {author} {\bibfnamefont {M.~B.}\ \bibnamefont {Jungfleisch}},\ }\bibfield  {title} {\bibinfo {title} {Emergent spin dynamics enabled by lattice interactions in a bicomponent artificial spin ice},\ }\href@noop {} {\bibfield  {journal} {\bibinfo  {journal} {Nano Letters}\ }\textbf {\bibinfo {volume} {21}},\ \bibinfo {pages} {1921} (\bibinfo {year} {2021}{\natexlab{a}})}\BibitemShut {NoStop}%
\bibitem [{\citenamefont {Ghosh}\ \emph {et~al.}(2019)\citenamefont {Ghosh}, \citenamefont {Ma}, \citenamefont {Lourembam}, \citenamefont {Jin}, \citenamefont {Maddu}, \citenamefont {Yap},\ and\ \citenamefont {Ter~Lim}}]{ghosh2019}%
  \BibitemOpen
  \bibfield  {author} {\bibinfo {author} {\bibfnamefont {A.}~\bibnamefont {Ghosh}}, \bibinfo {author} {\bibfnamefont {F.}~\bibnamefont {Ma}}, \bibinfo {author} {\bibfnamefont {J.}~\bibnamefont {Lourembam}}, \bibinfo {author} {\bibfnamefont {X.}~\bibnamefont {Jin}}, \bibinfo {author} {\bibfnamefont {R.}~\bibnamefont {Maddu}}, \bibinfo {author} {\bibfnamefont {Q.~J.}\ \bibnamefont {Yap}},\ and\ \bibinfo {author} {\bibfnamefont {S.}~\bibnamefont {Ter~Lim}},\ }\bibfield  {title} {\bibinfo {title} {Emergent dynamics of artificial spin-ice lattice based on an ultrathin ferromagnet},\ }\href@noop {} {\bibfield  {journal} {\bibinfo  {journal} {Nano Letters}\ }\textbf {\bibinfo {volume} {20}},\ \bibinfo {pages} {109} (\bibinfo {year} {2019})}\BibitemShut {NoStop}%
\bibitem [{\citenamefont {Talapatra}\ \emph {et~al.}(2020)\citenamefont {Talapatra}, \citenamefont {Singh},\ and\ \citenamefont {Adeyeye}}]{talapatra2020}%
  \BibitemOpen
  \bibfield  {author} {\bibinfo {author} {\bibfnamefont {A.}~\bibnamefont {Talapatra}}, \bibinfo {author} {\bibfnamefont {N.}~\bibnamefont {Singh}},\ and\ \bibinfo {author} {\bibfnamefont {A.~O.}\ \bibnamefont {Adeyeye}},\ }\bibfield  {title} {\bibinfo {title} {Magnetic tunability of permalloy artificial spin ice structures},\ }\href {https://doi.org/10.1103/PhysRevApplied.13.014034} {\bibfield  {journal} {\bibinfo  {journal} {Phys. Rev. Appl.}\ }\textbf {\bibinfo {volume} {13}},\ \bibinfo {pages} {014034} (\bibinfo {year} {2020})}\BibitemShut {NoStop}%
\bibitem [{\citenamefont {Gartside}\ \emph {et~al.}(2021)\citenamefont {Gartside}, \citenamefont {Vanstone}, \citenamefont {Dion}, \citenamefont {Stenning}, \citenamefont {Arroo}, \citenamefont {Kurebayashi},\ and\ \citenamefont {Branford}}]{gartside2021}%
  \BibitemOpen
  \bibfield  {author} {\bibinfo {author} {\bibfnamefont {J.~C.}\ \bibnamefont {Gartside}}, \bibinfo {author} {\bibfnamefont {A.}~\bibnamefont {Vanstone}}, \bibinfo {author} {\bibfnamefont {T.}~\bibnamefont {Dion}}, \bibinfo {author} {\bibfnamefont {K.~D.}\ \bibnamefont {Stenning}}, \bibinfo {author} {\bibfnamefont {D.~M.}\ \bibnamefont {Arroo}}, \bibinfo {author} {\bibfnamefont {H.}~\bibnamefont {Kurebayashi}},\ and\ \bibinfo {author} {\bibfnamefont {W.~R.}\ \bibnamefont {Branford}},\ }\bibfield  {title} {\bibinfo {title} {Reconfigurable magnonic mode-hybridisation and spectral control in a bicomponent artificial spin ice},\ }\href@noop {} {\bibfield  {journal} {\bibinfo  {journal} {Nature communications}\ }\textbf {\bibinfo {volume} {12}},\ \bibinfo {pages} {2488} (\bibinfo {year} {2021})}\BibitemShut {NoStop}%
\bibitem [{\citenamefont {Arora}\ and\ \citenamefont {Das}(2022)}]{arora2022}%
  \BibitemOpen
  \bibfield  {author} {\bibinfo {author} {\bibfnamefont {N.}~\bibnamefont {Arora}}\ and\ \bibinfo {author} {\bibfnamefont {P.}~\bibnamefont {Das}},\ }\bibfield  {title} {\bibinfo {title} {Excitation of spin waves in the presence of magnetic charges and monopole polarons in finite-size square artificial spin ice systems},\ }\href@noop {} {\bibfield  {journal} {\bibinfo  {journal} {Physical Review B}\ }\textbf {\bibinfo {volume} {106}},\ \bibinfo {pages} {184411} (\bibinfo {year} {2022})}\BibitemShut {NoStop}%
\bibitem [{\citenamefont {Kuchibhotla}\ \emph {et~al.}(2023)\citenamefont {Kuchibhotla}, \citenamefont {Haldar},\ and\ \citenamefont {Adeyeye}}]{kuchibhotla2023}%
  \BibitemOpen
  \bibfield  {author} {\bibinfo {author} {\bibfnamefont {M.}~\bibnamefont {Kuchibhotla}}, \bibinfo {author} {\bibfnamefont {A.}~\bibnamefont {Haldar}},\ and\ \bibinfo {author} {\bibfnamefont {A.~O.}\ \bibnamefont {Adeyeye}},\ }\bibfield  {title} {\bibinfo {title} {Field angle dependent resonant dynamics of artificial spin ice lattices},\ }\href@noop {} {\bibfield  {journal} {\bibinfo  {journal} {Nanotechnology}\ }\textbf {\bibinfo {volume} {34}},\ \bibinfo {pages} {325302} (\bibinfo {year} {2023})}\BibitemShut {NoStop}%
\bibitem [{\citenamefont {Gliga}\ \emph {et~al.}(2015)\citenamefont {Gliga}, \citenamefont {K\'akay}, \citenamefont {Heyderman}, \citenamefont {Hertel},\ and\ \citenamefont {Heinonen}}]{gliga2015}%
  \BibitemOpen
  \bibfield  {author} {\bibinfo {author} {\bibfnamefont {S.}~\bibnamefont {Gliga}}, \bibinfo {author} {\bibfnamefont {A.}~\bibnamefont {K\'akay}}, \bibinfo {author} {\bibfnamefont {L.~J.}\ \bibnamefont {Heyderman}}, \bibinfo {author} {\bibfnamefont {R.}~\bibnamefont {Hertel}},\ and\ \bibinfo {author} {\bibfnamefont {O.~G.}\ \bibnamefont {Heinonen}},\ }\bibfield  {title} {\bibinfo {title} {Broken vertex symmetry and finite zero-point entropy in the artificial square ice ground state},\ }\href {https://doi.org/10.1103/PhysRevB.92.060413} {\bibfield  {journal} {\bibinfo  {journal} {Phys. Rev. B}\ }\textbf {\bibinfo {volume} {92}},\ \bibinfo {pages} {060413} (\bibinfo {year} {2015})}\BibitemShut {NoStop}%
\bibitem [{\citenamefont {Heyderman}\ and\ \citenamefont {Stamps}(2013)}]{heyderman2013}%
  \BibitemOpen
  \bibfield  {author} {\bibinfo {author} {\bibfnamefont {L.~J.}\ \bibnamefont {Heyderman}}\ and\ \bibinfo {author} {\bibfnamefont {R.~L.}\ \bibnamefont {Stamps}},\ }\bibfield  {title} {\bibinfo {title} {Artificial ferroic systems: novel functionality from structure, interactions and dynamics},\ }\href@noop {} {\bibfield  {journal} {\bibinfo  {journal} {Journal of Physics: Condensed Matter}\ }\textbf {\bibinfo {volume} {25}},\ \bibinfo {pages} {363201} (\bibinfo {year} {2013})}\BibitemShut {NoStop}%
\bibitem [{\citenamefont {Iacocca}\ \emph {et~al.}(2020)\citenamefont {Iacocca}, \citenamefont {Gliga},\ and\ \citenamefont {Heinonen}}]{iacocca2020}%
  \BibitemOpen
  \bibfield  {author} {\bibinfo {author} {\bibfnamefont {E.}~\bibnamefont {Iacocca}}, \bibinfo {author} {\bibfnamefont {S.}~\bibnamefont {Gliga}},\ and\ \bibinfo {author} {\bibfnamefont {O.~G.}\ \bibnamefont {Heinonen}},\ }\bibfield  {title} {\bibinfo {title} {Tailoring spin-wave channels in a reconfigurable artificial spin ice},\ }\href@noop {} {\bibfield  {journal} {\bibinfo  {journal} {Physical Review Applied}\ }\textbf {\bibinfo {volume} {13}},\ \bibinfo {pages} {044047} (\bibinfo {year} {2020})}\BibitemShut {NoStop}%
\bibitem [{\citenamefont {Lendinez}\ \emph {et~al.}(2021{\natexlab{b}})\citenamefont {Lendinez}, \citenamefont {Taghipour~Kaffash},\ and\ \citenamefont {Jungfleisch}}]{lendinez2021obser}%
  \BibitemOpen
  \bibfield  {author} {\bibinfo {author} {\bibfnamefont {S.}~\bibnamefont {Lendinez}}, \bibinfo {author} {\bibfnamefont {M.}~\bibnamefont {Taghipour~Kaffash}},\ and\ \bibinfo {author} {\bibfnamefont {M.~B.}\ \bibnamefont {Jungfleisch}},\ }\bibfield  {title} {\bibinfo {title} {Observation of mode splitting in artificial spin ice: A comparative ferromagnetic resonance and brillouin light scattering study},\ }\href@noop {} {\bibfield  {journal} {\bibinfo  {journal} {Applied Physics Letters}\ }\textbf {\bibinfo {volume} {118}} (\bibinfo {year} {2021}{\natexlab{b}})}\BibitemShut {NoStop}%
\bibitem [{\citenamefont {Vansteenkiste}\ \emph {et~al.}(2014)\citenamefont {Vansteenkiste}, \citenamefont {Leliaert}, \citenamefont {Dvornik}, \citenamefont {Helsen}, \citenamefont {Garcia-Sanchez},\ and\ \citenamefont {Van~Waeyenberge}}]{vansteenkiste2014}%
  \BibitemOpen
  \bibfield  {author} {\bibinfo {author} {\bibfnamefont {A.}~\bibnamefont {Vansteenkiste}}, \bibinfo {author} {\bibfnamefont {J.}~\bibnamefont {Leliaert}}, \bibinfo {author} {\bibfnamefont {M.}~\bibnamefont {Dvornik}}, \bibinfo {author} {\bibfnamefont {M.}~\bibnamefont {Helsen}}, \bibinfo {author} {\bibfnamefont {F.}~\bibnamefont {Garcia-Sanchez}},\ and\ \bibinfo {author} {\bibfnamefont {B.}~\bibnamefont {Van~Waeyenberge}},\ }\bibfield  {title} {\bibinfo {title} {The design and verification of mumax3},\ }\href@noop {} {\bibfield  {journal} {\bibinfo  {journal} {AIP advances}\ }\textbf {\bibinfo {volume} {4}} (\bibinfo {year} {2014})}\BibitemShut {NoStop}%
\bibitem [{\citenamefont {Buschow}(2003)}]{buschow2003}%
  \BibitemOpen
  \bibfield  {author} {\bibinfo {author} {\bibfnamefont {K.~J.}\ \bibnamefont {Buschow}},\ }\href@noop {} {\emph {\bibinfo {title} {Handbook of magnetic materials}}}\ (\bibinfo  {publisher} {Elsevier},\ \bibinfo {year} {2003})\BibitemShut {NoStop}%
\bibitem [{\citenamefont {Wei}\ \emph {et~al.}(2016)\citenamefont {Wei}, \citenamefont {Zhu}, \citenamefont {Song}, \citenamefont {Feng}, \citenamefont {Jing}, \citenamefont {Wang}, \citenamefont {Liu},\ and\ \citenamefont {Wang}}]{wei2016}%
  \BibitemOpen
  \bibfield  {author} {\bibinfo {author} {\bibfnamefont {J.}~\bibnamefont {Wei}}, \bibinfo {author} {\bibfnamefont {Z.}~\bibnamefont {Zhu}}, \bibinfo {author} {\bibfnamefont {C.}~\bibnamefont {Song}}, \bibinfo {author} {\bibfnamefont {H.}~\bibnamefont {Feng}}, \bibinfo {author} {\bibfnamefont {P.}~\bibnamefont {Jing}}, \bibinfo {author} {\bibfnamefont {X.}~\bibnamefont {Wang}}, \bibinfo {author} {\bibfnamefont {Q.}~\bibnamefont {Liu}},\ and\ \bibinfo {author} {\bibfnamefont {J.}~\bibnamefont {Wang}},\ }\bibfield  {title} {\bibinfo {title} {Annealing influence on the exchange stiffness constant of permalloy films with stripe domains},\ }\href@noop {} {\bibfield  {journal} {\bibinfo  {journal} {Journal of Physics D: Applied Physics}\ }\textbf {\bibinfo {volume} {49}},\ \bibinfo {pages} {265002} (\bibinfo {year} {2016})}\BibitemShut {NoStop}%
\bibitem [{\citenamefont {Stoner}\ and\ \citenamefont {Wohlfarth}(1948)}]{stoner1948}%
  \BibitemOpen
  \bibfield  {author} {\bibinfo {author} {\bibfnamefont {E.~C.}\ \bibnamefont {Stoner}}\ and\ \bibinfo {author} {\bibfnamefont {E.}~\bibnamefont {Wohlfarth}},\ }\bibfield  {title} {\bibinfo {title} {A mechanism of magnetic hysteresis in heterogeneous alloys},\ }\href@noop {} {\bibfield  {journal} {\bibinfo  {journal} {Philosophical Transactions of the Royal Society of London. Series A, Mathematical and Physical Sciences}\ }\textbf {\bibinfo {volume} {240}},\ \bibinfo {pages} {599} (\bibinfo {year} {1948})}\BibitemShut {NoStop}%
\bibitem [{\citenamefont {Coey}(2010)}]{coey2010}%
  \BibitemOpen
  \bibfield  {author} {\bibinfo {author} {\bibfnamefont {J.~M.}\ \bibnamefont {Coey}},\ }\href@noop {} {\emph {\bibinfo {title} {Magnetism and magnetic materials}}}\ (\bibinfo  {publisher} {Cambridge university press},\ \bibinfo {year} {2010})\BibitemShut {NoStop}%
\bibitem [{\citenamefont {Guimar{\~a}es}\ and\ \citenamefont {Guimaraes}(2009)}]{guimaraes2009}%
  \BibitemOpen
  \bibfield  {author} {\bibinfo {author} {\bibfnamefont {A.~P.}\ \bibnamefont {Guimar{\~a}es}}\ and\ \bibinfo {author} {\bibfnamefont {A.~P.}\ \bibnamefont {Guimaraes}},\ }\href@noop {} {\emph {\bibinfo {title} {Principles of nanomagnetism}}},\ Vol.~\bibinfo {volume} {7}\ (\bibinfo  {publisher} {Springer},\ \bibinfo {year} {2009})\BibitemShut {NoStop}%
\bibitem [{\citenamefont {Kittel}(1947)}]{kittel1947}%
  \BibitemOpen
  \bibfield  {author} {\bibinfo {author} {\bibfnamefont {C.}~\bibnamefont {Kittel}},\ }\bibfield  {title} {\bibinfo {title} {Interpretation of anomalous larmor frequencies in ferromagnetic resonance experiment},\ }\href@noop {} {\bibfield  {journal} {\bibinfo  {journal} {Physical Review}\ }\textbf {\bibinfo {volume} {71}},\ \bibinfo {pages} {270} (\bibinfo {year} {1947})}\BibitemShut {NoStop}%
\bibitem [{\citenamefont {McMichael}\ and\ \citenamefont {Maranville}(2006)}]{McMichael2006}%
  \BibitemOpen
  \bibfield  {author} {\bibinfo {author} {\bibfnamefont {R.~D.}\ \bibnamefont {McMichael}}\ and\ \bibinfo {author} {\bibfnamefont {B.~B.}\ \bibnamefont {Maranville}},\ }\bibfield  {title} {\bibinfo {title} {Edge saturation fields and dynamic edge modes in ideal and nonideal magnetic film edges},\ }\href@noop {} {\bibfield  {journal} {\bibinfo  {journal} {Phys. Rev. B}\ }\textbf {\bibinfo {volume} {74}},\ \bibinfo {pages} {024424} (\bibinfo {year} {2006})}\BibitemShut {NoStop}%
\bibitem [{\citenamefont {Goryca}\ \emph {et~al.}(2022)\citenamefont {Goryca}, \citenamefont {Zhang}, \citenamefont {Watts}, \citenamefont {Nisoli}, \citenamefont {Leighton}, \citenamefont {Schiffer},\ and\ \citenamefont {Crooker}}]{goryca2022}%
  \BibitemOpen
  \bibfield  {author} {\bibinfo {author} {\bibfnamefont {M.}~\bibnamefont {Goryca}}, \bibinfo {author} {\bibfnamefont {X.}~\bibnamefont {Zhang}}, \bibinfo {author} {\bibfnamefont {J.~D.}\ \bibnamefont {Watts}}, \bibinfo {author} {\bibfnamefont {C.}~\bibnamefont {Nisoli}}, \bibinfo {author} {\bibfnamefont {C.}~\bibnamefont {Leighton}}, \bibinfo {author} {\bibfnamefont {P.}~\bibnamefont {Schiffer}},\ and\ \bibinfo {author} {\bibfnamefont {S.~A.}\ \bibnamefont {Crooker}},\ }\bibfield  {title} {\bibinfo {title} {Magnetic field dependent thermodynamic properties of square and quadrupolar artificial spin ice},\ }\href {https://doi.org/10.1103/PhysRevB.105.094406} {\bibfield  {journal} {\bibinfo  {journal} {Phys. Rev. B}\ }\textbf {\bibinfo {volume} {105}},\ \bibinfo {pages} {094406} (\bibinfo {year} {2022})}\BibitemShut {NoStop}%
\bibitem [{\citenamefont {Porro}\ \emph {et~al.}(2019)\citenamefont {Porro}, \citenamefont {Morley}, \citenamefont {Venero}, \citenamefont {Mac{\^e}do}, \citenamefont {Rosamond}, \citenamefont {Linfield}, \citenamefont {Stamps}, \citenamefont {Marrows},\ and\ \citenamefont {Langridge}}]{porro2019}%
  \BibitemOpen
  \bibfield  {author} {\bibinfo {author} {\bibfnamefont {J.~M.}\ \bibnamefont {Porro}}, \bibinfo {author} {\bibfnamefont {S.~A.}\ \bibnamefont {Morley}}, \bibinfo {author} {\bibfnamefont {D.~A.}\ \bibnamefont {Venero}}, \bibinfo {author} {\bibfnamefont {R.}~\bibnamefont {Mac{\^e}do}}, \bibinfo {author} {\bibfnamefont {M.~C.}\ \bibnamefont {Rosamond}}, \bibinfo {author} {\bibfnamefont {E.~H.}\ \bibnamefont {Linfield}}, \bibinfo {author} {\bibfnamefont {R.~L.}\ \bibnamefont {Stamps}}, \bibinfo {author} {\bibfnamefont {C.~H.}\ \bibnamefont {Marrows}},\ and\ \bibinfo {author} {\bibfnamefont {S.}~\bibnamefont {Langridge}},\ }\bibfield  {title} {\bibinfo {title} {Magnetization dynamics of weakly interacting sub-100 nm square artificial spin ices},\ }\href@noop {} {\bibfield  {journal} {\bibinfo  {journal} {Scientific reports}\ }\textbf {\bibinfo {volume} {9}},\ \bibinfo {pages} {19967} (\bibinfo {year} {2019})}\BibitemShut {NoStop}%
\bibitem [{\citenamefont {M{\"o}ller}\ and\ \citenamefont {Moessner}(2006)}]{moller2006}%
  \BibitemOpen
  \bibfield  {author} {\bibinfo {author} {\bibfnamefont {G.}~\bibnamefont {M{\"o}ller}}\ and\ \bibinfo {author} {\bibfnamefont {R.}~\bibnamefont {Moessner}},\ }\bibfield  {title} {\bibinfo {title} {Artificial square ice and related dipolar nanoarrays},\ }\href@noop {} {\bibfield  {journal} {\bibinfo  {journal} {Physical Review Letters}\ }\textbf {\bibinfo {volume} {96}},\ \bibinfo {pages} {237202} (\bibinfo {year} {2006})}\BibitemShut {NoStop}%
\bibitem [{\citenamefont {Keswani}\ \emph {et~al.}(2021)\citenamefont {Keswani}, \citenamefont {Lopes}, \citenamefont {Nakajima}, \citenamefont {Singh}, \citenamefont {Chauhan}, \citenamefont {Som}, \citenamefont {Kumar}, \citenamefont {Pereira},\ and\ \citenamefont {Das}}]{keswani2021}%
  \BibitemOpen
  \bibfield  {author} {\bibinfo {author} {\bibfnamefont {N.}~\bibnamefont {Keswani}}, \bibinfo {author} {\bibfnamefont {R.~J.~C.}\ \bibnamefont {Lopes}}, \bibinfo {author} {\bibfnamefont {Y.}~\bibnamefont {Nakajima}}, \bibinfo {author} {\bibfnamefont {R.}~\bibnamefont {Singh}}, \bibinfo {author} {\bibfnamefont {N.}~\bibnamefont {Chauhan}}, \bibinfo {author} {\bibfnamefont {T.}~\bibnamefont {Som}}, \bibinfo {author} {\bibfnamefont {D.~S.}\ \bibnamefont {Kumar}}, \bibinfo {author} {\bibfnamefont {A.~R.}\ \bibnamefont {Pereira}},\ and\ \bibinfo {author} {\bibfnamefont {P.}~\bibnamefont {Das}},\ }\bibfield  {title} {\bibinfo {title} {Controlled creation and annihilation of isolated robust emergent magnetic monopole like charged vertices in square artificial spin ice},\ }\href@noop {} {\bibfield  {journal} {\bibinfo  {journal} {Scientific Reports}\ }\textbf {\bibinfo {volume} {11}},\ \bibinfo {pages} {13593} (\bibinfo {year} {2021})}\BibitemShut {NoStop}%
\bibitem [{\citenamefont {Morley}\ \emph {et~al.}(2019)\citenamefont {Morley}, \citenamefont {Porro}, \citenamefont {Hrabec}, \citenamefont {Rosamond}, \citenamefont {Venero}, \citenamefont {Linfield}, \citenamefont {Burnell}, \citenamefont {Im}, \citenamefont {Fischer}, \citenamefont {Langridge},\ and\ \citenamefont {Marrows}}]{morley2019}%
  \BibitemOpen
  \bibfield  {author} {\bibinfo {author} {\bibfnamefont {S.~A.}\ \bibnamefont {Morley}}, \bibinfo {author} {\bibfnamefont {J.~M.}\ \bibnamefont {Porro}}, \bibinfo {author} {\bibfnamefont {A.}~\bibnamefont {Hrabec}}, \bibinfo {author} {\bibfnamefont {M.~C.}\ \bibnamefont {Rosamond}}, \bibinfo {author} {\bibfnamefont {D.~A.}\ \bibnamefont {Venero}}, \bibinfo {author} {\bibfnamefont {E.~H.}\ \bibnamefont {Linfield}}, \bibinfo {author} {\bibfnamefont {G.}~\bibnamefont {Burnell}}, \bibinfo {author} {\bibfnamefont {M.-Y.}\ \bibnamefont {Im}}, \bibinfo {author} {\bibfnamefont {P.}~\bibnamefont {Fischer}}, \bibinfo {author} {\bibfnamefont {S.}~\bibnamefont {Langridge}},\ and\ \bibinfo {author} {\bibfnamefont {C.~H.}\ \bibnamefont {Marrows}},\ }\bibfield  {title} {\bibinfo {title} {Thermally and field-driven mobility of emergent magnetic charges in square artificial spin ice},\ }\href@noop {} {\bibfield  {journal} {\bibinfo  {journal} {Scientific reports}\ }\textbf {\bibinfo {volume} {9}},\ \bibinfo {pages} {15989}
  (\bibinfo {year} {2019})}\BibitemShut {NoStop}%
\bibitem [{\citenamefont {Zhang}\ \emph {et~al.}(1994)\citenamefont {Zhang}, \citenamefont {Zhou}, \citenamefont {Wigen},\ and\ \citenamefont {Ounadjela}}]{Zhang1994}%
  \BibitemOpen
  \bibfield  {author} {\bibinfo {author} {\bibfnamefont {Z.}~\bibnamefont {Zhang}}, \bibinfo {author} {\bibfnamefont {L.}~\bibnamefont {Zhou}}, \bibinfo {author} {\bibfnamefont {P.~E.}\ \bibnamefont {Wigen}},\ and\ \bibinfo {author} {\bibfnamefont {K.}~\bibnamefont {Ounadjela}},\ }\bibfield  {title} {\bibinfo {title} {Using ferromagnetic resonance as a sensitive method to study temperature dependence of interlayer exchange coupling},\ }\href@noop {} {\bibfield  {journal} {\bibinfo  {journal} {Phys. Rev. Lett.}\ }\textbf {\bibinfo {volume} {73}},\ \bibinfo {pages} {336} (\bibinfo {year} {1994})}\BibitemShut {NoStop}%
\bibitem [{\citenamefont {Belmeguenai}\ \emph {et~al.}(2008)\citenamefont {Belmeguenai}, \citenamefont {Martin}, \citenamefont {Woltersdorf}, \citenamefont {Bayreuther}, \citenamefont {Baltz}, \citenamefont {Suszka},\ and\ \citenamefont {Hickey}}]{Belmeguenai2008}%
  \BibitemOpen
  \bibfield  {author} {\bibinfo {author} {\bibfnamefont {M.}~\bibnamefont {Belmeguenai}}, \bibinfo {author} {\bibfnamefont {T.}~\bibnamefont {Martin}}, \bibinfo {author} {\bibfnamefont {G.}~\bibnamefont {Woltersdorf}}, \bibinfo {author} {\bibfnamefont {G.}~\bibnamefont {Bayreuther}}, \bibinfo {author} {\bibfnamefont {V.}~\bibnamefont {Baltz}}, \bibinfo {author} {\bibfnamefont {A.~K.}\ \bibnamefont {Suszka}},\ and\ \bibinfo {author} {\bibfnamefont {B.~J.}\ \bibnamefont {Hickey}},\ }\bibfield  {title} {\bibinfo {title} {Microwave spectroscopy with vector network analyzer for interlayer exchange-coupled symmetrical and asymmetrical {NiFe}/{Ru}/{NiFe}},\ }\href@noop {} {\bibfield  {journal} {\bibinfo  {journal} {Journal of Physics: Condensed Matter}\ }\textbf {\bibinfo {volume} {20}},\ \bibinfo {pages} {345206} (\bibinfo {year} {2008})}\BibitemShut {NoStop}%
\bibitem [{\citenamefont {Gonzalez-Chavez}\ \emph {et~al.}(2013)\citenamefont {Gonzalez-Chavez}, \citenamefont {Dutra}, \citenamefont {Rosa}, \citenamefont {Marcondes}, \citenamefont {Mello},\ and\ \citenamefont {Sommer}}]{gonzalez2013}%
  \BibitemOpen
  \bibfield  {author} {\bibinfo {author} {\bibfnamefont {D.~E.}\ \bibnamefont {Gonzalez-Chavez}}, \bibinfo {author} {\bibfnamefont {R.}~\bibnamefont {Dutra}}, \bibinfo {author} {\bibfnamefont {W.~O.}\ \bibnamefont {Rosa}}, \bibinfo {author} {\bibfnamefont {T.~L.}\ \bibnamefont {Marcondes}}, \bibinfo {author} {\bibfnamefont {A.}~\bibnamefont {Mello}},\ and\ \bibinfo {author} {\bibfnamefont {R.~L.}\ \bibnamefont {Sommer}},\ }\bibfield  {title} {\bibinfo {title} {Interlayer coupling in spin valves studied by broadband ferromagnetic resonance},\ }\href {https://doi.org/10.1103/PhysRevB.88.104431} {\bibfield  {journal} {\bibinfo  {journal} {Phys. Rev. B}\ }\textbf {\bibinfo {volume} {88}},\ \bibinfo {pages} {104431} (\bibinfo {year} {2013})}\BibitemShut {NoStop}%
\bibitem [{\citenamefont {Pervez}\ \emph {et~al.}(2022)\citenamefont {Pervez}, \citenamefont {Gonzalez-Chavez}, \citenamefont {Dutra}, \citenamefont {Silva}, \citenamefont {Raza},\ and\ \citenamefont {Sommer}}]{Pervez2022}%
  \BibitemOpen
  \bibfield  {author} {\bibinfo {author} {\bibfnamefont {M.~A.}\ \bibnamefont {Pervez}}, \bibinfo {author} {\bibfnamefont {D.}~\bibnamefont {Gonzalez-Chavez}}, \bibinfo {author} {\bibfnamefont {R.}~\bibnamefont {Dutra}}, \bibinfo {author} {\bibfnamefont {B.}~\bibnamefont {Silva}}, \bibinfo {author} {\bibfnamefont {S.}~\bibnamefont {Raza}},\ and\ \bibinfo {author} {\bibfnamefont {R.}~\bibnamefont {Sommer}},\ }\bibfield  {title} {\bibinfo {title} {Damping in synthetic antiferromagnets},\ }\href@noop {} {\bibfield  {journal} {\bibinfo  {journal} {Journal of Magnetism and Magnetic Materials}\ }\textbf {\bibinfo {volume} {548}},\ \bibinfo {pages} {168923} (\bibinfo {year} {2022})}\BibitemShut {NoStop}%
\end{thebibliography}%

\end{document}